\documentclass[a4paper,fleqn,usenatbib]{mn2e}

\usepackage[utf8]{inputenc}
\usepackage[T1]{fontenc}
\usepackage{lmodern} 

\usepackage[T1]{fontenc}
\usepackage{ae,aecompl}

\usepackage{graphicx}   
\usepackage{amsmath}    
\usepackage{amssymb}    

\newcommand{\unit}[1]{\, \rm{#1}}	
\newcommand{\sscript}[1]{{_{\rm{#1}}}}		

\title[A semi-analytic dynamical friction model for cored galaxies]{A semi-analytic dynamical friction model for cored galaxies}
\author[J. A. Petts, J. I. Read,A. Gualandris]{J. A. Petts$^{1}$\thanks{E-mail:
j.petts@surrey.ac.uk}, J. I. Read, A. Gualandris \\
$^{1}$University of Surrey, Department of Physics, Guildford, United Kingdom}
\begin{document}

\date{TEMP}

\pagerange{\pageref{firstpage}--\pageref{lastpage}} \pubyear{2016}

\maketitle

\label{firstpage}

\begin{abstract}
	We present a dynamical friction model based on Chandrasekhar's formula that reproduces the fast inspiral and stalling experienced by satellites orbiting galaxies with a large constant density core. We show that the fast inspiral phase does not owe to resonance. Rather, it owes to the background velocity distribution function for the constant density core being dissimilar from the usually-assumed Maxwellian distribution. Using the correct background velocity distribution function and the semi-analytic model from \citet{Petts15}, we are able to correctly reproduce the infall rate in both cored and cusped potentials. However, in the case of large cores, our model is no longer able to correctly capture core-stalling. We show that this stalling owes to the tidal radius of the satellite approaching the size of the core. By switching off dynamical friction when $r\sscript{t}(r) = r$ (where $r\sscript{t}$ is the tidal radius at the satellite's position) we arrive at a model which reproduces the $N$-body results remarkably well. Since the tidal radius can be very large for constant density background distributions, our model recovers the result that stalling can occur for $M\sscript{s}/M\sscript{enc} \ll 1$, where $M\sscript{s}$ and $M\sscript{enc}$ are the mass of the satellite and the enclosed galaxy mass, respectively. Finally, we include the contribution to dynamical friction that comes from stars moving faster than the satellite. This next-to-leading order effect becomes the dominant driver of inspiral near the core region, prior to stalling.
\end{abstract}


\section{Introduction}
Dynamical friction is a drag force caused by momentum exchange between a massive object moving within a sea of lighter background `stars'. (We shall refer to these as `stars' throughout this paper, though they could comprise any self-gravitating entity; e.g. dark matter particles \citep{BT08}.) Dynamical friction is thought to be a key driver of the mergers of star clusters, galaxies, and even galaxy clusters over cosmic time \citep{Gan10,Peirani10,ArcaSedda14c,ArcaSedda15,Priyatikanto16}.

In a seminal work \citet{Chandrasekhar43} calculated the force on a massive object traversing an infinite homogeneous isotropic background. Approximating the drag to come only from stars moving slower than the satellite and ignoring the velocity dependence of the strength of interactions, Chandrasekhar's formula is often seen in the following simplified analytic form \citep{Chandrasekhar43, BT08}:

\begin{equation}
	\frac{d\bmath{v\sscript{S}}}{dt} = -4\pi G^2 M\sscript{S} \rho \log(\Lambda)f(v\sscript{*}<v\sscript{s})\frac{\bmath{v\sscript{S}}}{v\sscript{S}^3},
	\label{dynfric.eq}
\end{equation}
where $\log(\Lambda)$ is the Coulomb logarithm equal to $ \log{(b\sscript{max}/b\sscript{min})}$ where $b\sscript{max}$ and $b\sscript{min}$ are the maximum and minimum impact parameters for encounters, and it is assumed that $b\sscript{max} \gg b\sscript{min}$. $M\sscript{S}$ and $\bmath{v\sscript{S}}$ are the satellite's mass and velocity ($v\sscript{S} = \bmath{|v\sscript{S}|}$), $\rho$ is the background density and $f(v\sscript{*}<v\sscript{s})$ is the fraction of stars moving slower than the satellite.

Although derived under very simple assumptions, Chandrasekhar's formula has proven to be remarkably successful for more general isotropic spherical distributions. Chandrasekhar's formula has been shown to be a good approximation for low mass satellites due to the fact that the global response from the background appears to be negligible \citep{White83,Bontekoe87,Zaritsky88,Cora97}. For satellites more massive than $\gtrsim 10\%$ of the host galaxy mass, the global response is non-negligible. As such equation \ref{dynfric.eq} is not accurate for major mergers. 

However, for large mass ratios (e.g. dwarf galaxies moving in the halo of larger hosts, globular clusters in dwarf spheroidals, etc.), the formula has been very successful at reproducing the orbital evolution. Although the formula has remained largely unchanged, its application has seen many improvements in recent years \citep{Hashimoto03,Just05,Just11,Petts15}.

Typically, $\log(\Lambda)$ is poorly defined, with $b\sscript{max}$ being of the order of the size of the system and $b\sscript{min}$ the impact parameter for a 90 degree deflection \citep{BT08}. Because the impact parameters are found inside a logarithm which is slowly varying (so long as $b\sscript{max} \gg b\sscript{min}$), a constant $\log(\Lambda)$ is usually assumed, such that:

\begin{equation}
	\log(\Lambda) \sim \log\left(\frac{r\sscript{galaxy}}{\max(r\sscript{hm},\frac{GM\sscript{s}}{v^2\sscript{typ}})}\right),
\end{equation}
where $r\sscript{galaxy}$ is the effective ``size'' of the galaxy, and the minimum impact parameter is the larger of the size of the satellite and the impact parameter for which a star interacting at a ``typical velocity'', $v\sscript{typ}$, is deflected by 90 degrees \citep{BT08}.

\citet{Just11} showed that the approximation of local homogeneity is satisfied when $b\sscript{max}$ is set to be the length scale over which the density can be assumed to be constant \citep[see also][]{Just05}:

\begin{equation}
	b\sscript{max}(r) = \rm{min}\left(\frac{\rho(r)}{|\nabla\rho|},r\right),
\end{equation}
where $r$ is the galactocentric distance to the satellite, i.e. $b\sscript{max}$ varies with galactocentric distance and the slope of the profile \citep{Just05,Just11}. However, this prescription neglects the force from particles deep in the cuspy region of the galaxy, and under-predicts the drag when the satellite is very close to the centre. \citep[See][for a method for dealing with steep galactic centres.]{ArcaSedda14}

A radially varying $\log(\Lambda)$ improves the agreement with $N-$body models of the inspiral of eccentric satellites, where the satellite experiences a large variation in its radial position over an orbital time \citep[e.g.][]{Hashimoto03}. The variations of $\log(\Lambda)$ are also especially important when $b\sscript{max}/b\sscript{min}$ approaches unity \citep{Petts15}, which for most systems occurs at approximately $M\sscript{s} \sim M\sscript{enc}$ where the satellite is assumed to form  a quasi two-body system with the galactic centre and dynamical friction becomes inefficient \citep[e.g.][]{BT08, Gualandris08}. For galaxies with a large core, the inspiral of satellites stalls further out, at the edge of the constant density core \citep{Read06, Goerdt06,Inoue09,Goerdt10,Cole12}.

In \citet{Petts15} we introduced a semi-analytic model that reproduces the inspiral and correct stalling radii of satellites orbiting a Dehnen background \citep{Dehnen93}, including the case where the asymptotic logarithmic slope approached zero. However, the model fared less well for large constant density cores. It also failed to reproduce the rapid ``super-Chandrasekhar'' phase reported in \citet{Read06} (hereafter R06).

In this paper we show that this dramatic ``core-stalling'' effect can be approximately captured if we consider the radius at which the satellite tidally disrupts the core and sculpts the velocity distribution in this region. This idea was already explored in \citet{Goerdt10} for massive satellites infalling within relatively cuspy background distributions. Here, we show that this same idea can be generalised to large constant density cores in which the tidal radius of the satellite approaches the size of the cored region. We also address the ``Super-Chandrasekhar'' friction phase observed in R06 previously thought to be ``super-resonance'' of the harmonic core. The idea was that as the angular frequency in a perfectly flat core is the same for every star, some global resonant effects may drive the friction in a way that cannot be correctly described by considering only two body interactions (R06). As real systems are never truly harmonic -- especially if one considers the back reaction of the satellite -- we argue here that the friction cannot owe to ``super-resonance'' (i.e. a proposed efficient resonant interaction that dominates the friction, see R06). Instead, we show that this phase of rapid infall is due to previously invalid assumptions about the velocity distribution in the core, and can be explained entirely through local friction via two-body interactions.

The paper is organised as follows. In section \S \ref{models.ch} we describe the galaxy models used in this study. In section \S \ref{theory.ch} we explain the theory and necessary improvements to our model. In section \S \ref{simulations.ch} we describe the simulations used to test our model. In section \S \ref{results.ch} we compare the results of our new model to $N-$body results. In section \S \ref{discussion.ch} we discuss the stalling mechanism and the potentially related problem of ``dynamical buoyancy'' reported recently in \citet{Cole12}. Finally, in \S \ref{conclusion.ch}, we present our conclusions.

\section{Models}
\label{models.ch}

In this paper, we primarily consider the inspiral of a massive body moving in an isotropic distribution of stars described by Hénon's Isochrone model \citep{Henon59,Henon60}:

\begin{equation}
	\rho(r) = M\sscript{g} \left[ \frac{3(b+a)a^2 - r^2 (b+3a)}{4 \pi (b+a)^3 a^3}\right],
	\label{Henon_rho.eq}
\end{equation}
where $M\sscript{g}$ is the total galaxy mass, $b$ is the scale radius and $a= \sqrt{b^2 + r^2}$. This background distribution has a particularly large and flat constant density core that leads to dynamical friction stalling much further out than predicted in the semi-analytic model from \citet{Petts15}.

\begin{figure}
 \includegraphics[width=\linewidth]{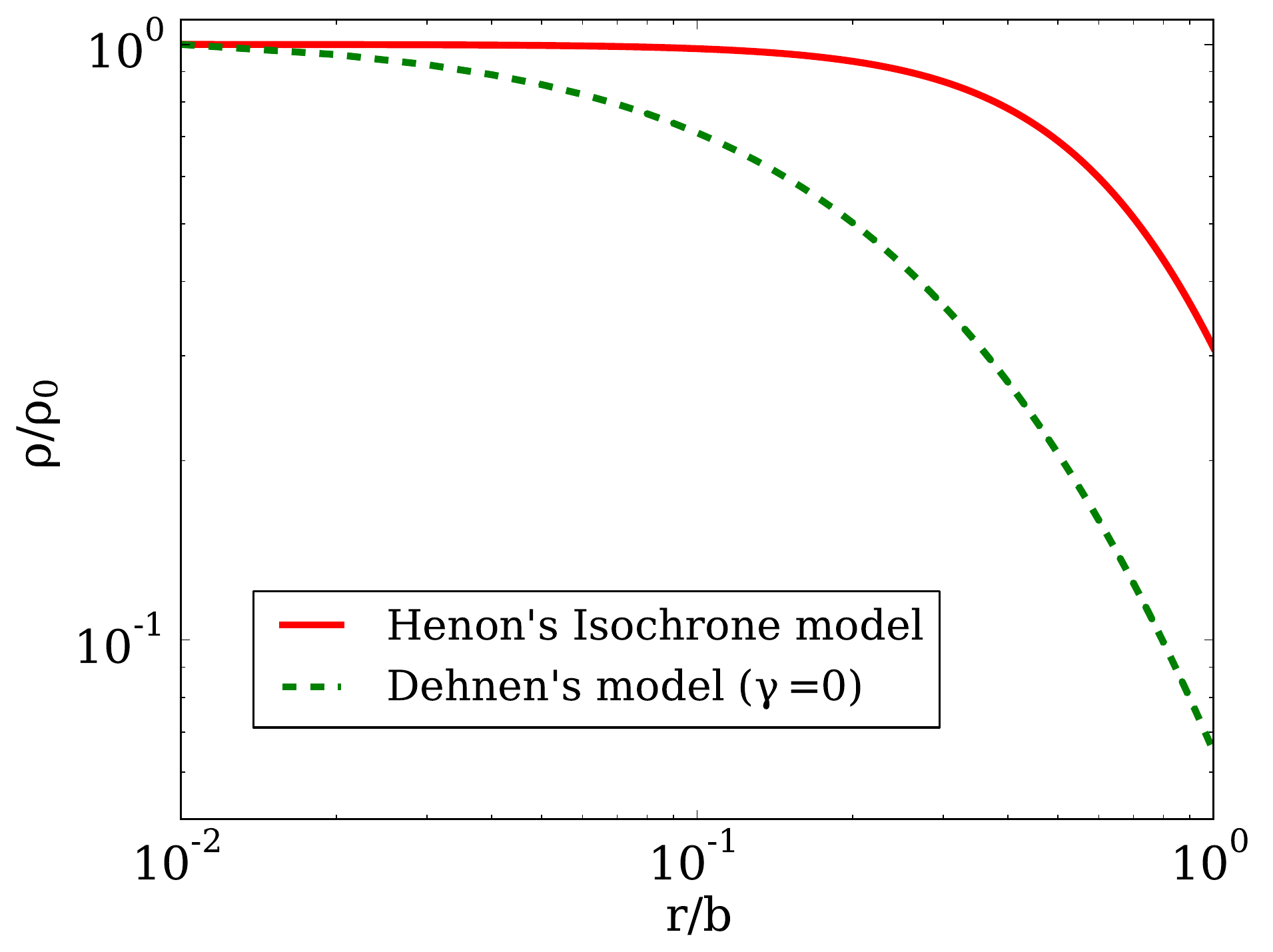}
 \caption{Density profiles of Hénon's Isochrone (solid red line), and Dehnen's model with $\gamma = 0$ (dashed green line); $r$ is normalised in units of $b$, and $\rho$ in units of and $\rho\sscript{0}$, the central density.} 
 \label{density.fig}
\end{figure}

In addition, in order to understand the ``super-Chandrasekhar'' phase that precedes stalling in large cores like the Hénon model, above, we consider also an isotropic Dehnen model background \citep{Dehnen93,Saha93,Tremaine94}:

\begin{equation}
	\rho(r) = \frac{(3-\gamma)M\sscript{g}}{4 \pi} \frac{b}{r^\gamma (r + b)^{4-\gamma}},
	\label{Dehnen_rho.eq}
\end{equation}
where $M\sscript{g}$ is the total galaxy mass, $b$ is the scale radius and $\gamma$ is the central log-slope of the model. We consider a model with $\gamma=0$, in which satellites in \citet{Petts15} exhibited the ``super Chandrasekhar'' phase but did not exhibit stalling at $M\sscript{s} \gg M\sscript{enc}$; and a cuspy model with $\gamma = 1.0$ which shows neither a ``super Chandrasekhar'' phase nor unexpected stalling. The Dehnen model, for both $\gamma=0$ and $\gamma=1.0$, is very well fit by the semi-analytic dynamical friction model from \citet{Petts15}. The Hénon Isochrone and Dehnen model background density distributions are shown in Fig. \ref{density.fig}.

\section{Theory}
\label{theory.ch}

\subsection{Semi-analytical dynamical friction model}
\label{Pett15.ch}

In \citet{Petts15} we presented a semi-analytical dynamical friction model based on equation \ref{dynfric.eq} with a Coulomb logarithm that varies as a function of satellite position and velocity:

\begin{equation}
	\log(\Lambda) = \log\left(\frac{b\sscript{max}}{b\sscript{min}}\right) = \log\left(\frac{\mathrm{min}(\rho(r)/|\nabla\rho(r)|,r)}{\mathrm{max}\left(r\sscript{hm}, GM\sscript{s}/v\sscript{s}^2\right)}\right),
	\label{coulog.eq}
\end{equation}
where $r$ is the satellite's galactocentric distance, $\rho(r)$ and $\nabla\rho(r)$ are the local density and first derivative of density at $r$, respectively. The maximum impact parameter is the region over which the density is approximately constant, so that the use of the local density, $\rho(r)$, is justified. If this exceeds the galactocentric distance, the galactocentric distance is used instead \citep[see][for more information]{Just11}. The minimum impact parameter is given by the maximum of satellite size and the impact parameter for a 90 degree deflection from a typical interaction. In \citet{Petts15} we considered that in an isotropic distribution some interactions occur at $v\sscript{rel}\sim 0$ and some as fast as  $v\sscript{rel}\sim 2v\sscript{s}$ (where $v\sscript{rel}$ is the relative velocity), so that on average the speed of the background particle is negligible, and the mean is $v\sscript{typ} \approx v\sscript{s}$. This is verified in section \S \ref{impact_parameters.ch}.

Our variable $\log(\Lambda)$ meant our model reproduced the satellite stalling that occurs in cuspy profiles, whereby the satellite stalls when it approximately encloses its own mass:

\begin{subequations}
\begin{align}
v\sscript{typ}^2 &\sim \frac{GM\sscript{g}(r)}{r} \sim \frac{GM\sscript{g}(r)}{b\sscript{max}},\\
b\sscript{min}&\sim \frac{GM\sscript{s}}{v\sscript{typ}^2} \sim \frac{M\sscript{s}}{M\sscript{g}(r)}b\sscript{max},\\
\frac{b\sscript{min}}{b\sscript{max}} &\sim \frac{M\sscript{s}}{M\sscript{g}(r)},
\end{align}
\label{stalling.eq}
\end{subequations}
where $M\sscript{g}(r)$ is the galaxy mass enclosed within $r$. This model reproduces both the inspiral and stalling radius for Dehnen's model for all values of the central slope, $\gamma$, excellently. However, in the errata of \citet{Petts15}, we show that this model does very badly for galaxies with a large constant density core, where it under-predicts both the frictional force and the stalling radius. In section \S \ref{fv.ch} we show that the inspiral can be well reproduced by relaxing the assumption of a Maxwellian velocity distribution, and in section \S \ref{core_stalling.ch} we show that the satellite stalls due to the large tidal radius of the satellite in the constant density region.

\subsection{A more accurate formula when $\mathbf{b\sscript{min} \sim b\sscript{max}}$}
\label{lambda_treatment.ch}

When studying the entire inspiral of satellites using Chandrasekhar's formalism, $b\sscript{min}$ can approach (or exceed in distributions without a large core) $b\sscript{max}$. It is therefore necessary to relax the assumption that $b\sscript{max} \gg b\sscript{min}$. Neglecting the velocity dependence of equation 27 from \citet{Chandrasekhar43}, equation \ref{dynfric.eq} originates from assuming $\Lambda \gg 1$:

\begin{align}
	\log(\Lambda^2 + 1) \simeq \log(\Lambda^2) = 2\log(\Lambda).
\end{align}
The factor of 2 is included in the coefficient of equation \ref{dynfric.eq}, such that if the approximation is not made then equation \ref{dynfric.eq} becomes:

\begin{equation}
	\frac{d\bmath{v\sscript{S}}}{dt} = -2\pi G^2 M\sscript{S} \rho \log(\Lambda^2 + 1)f(v\sscript{*}<v\sscript{s})\frac{\bmath{v\sscript{S}}}{v\sscript{S}^3},
	\label{dynfric2.eq}
\end{equation}
Therefore when $\Lambda < 1$, $\Lambda^2 \ll 1$ and the logarithm sharply tends to 0. In \citet{Petts15} we used equation \ref{dynfric.eq} and simply set $\log(\Lambda)$ to 0 if $\Lambda \leq 1$. The quantitative difference of the two approaches is minor, but equation \ref{dynfric2.eq} is more elegant, with no arbitrary cutoff.

\subsection{Reproducing the inspiral: the importance of velocity structure}
\label{fv.ch}

\begin{figure}
 \includegraphics[width=\linewidth]{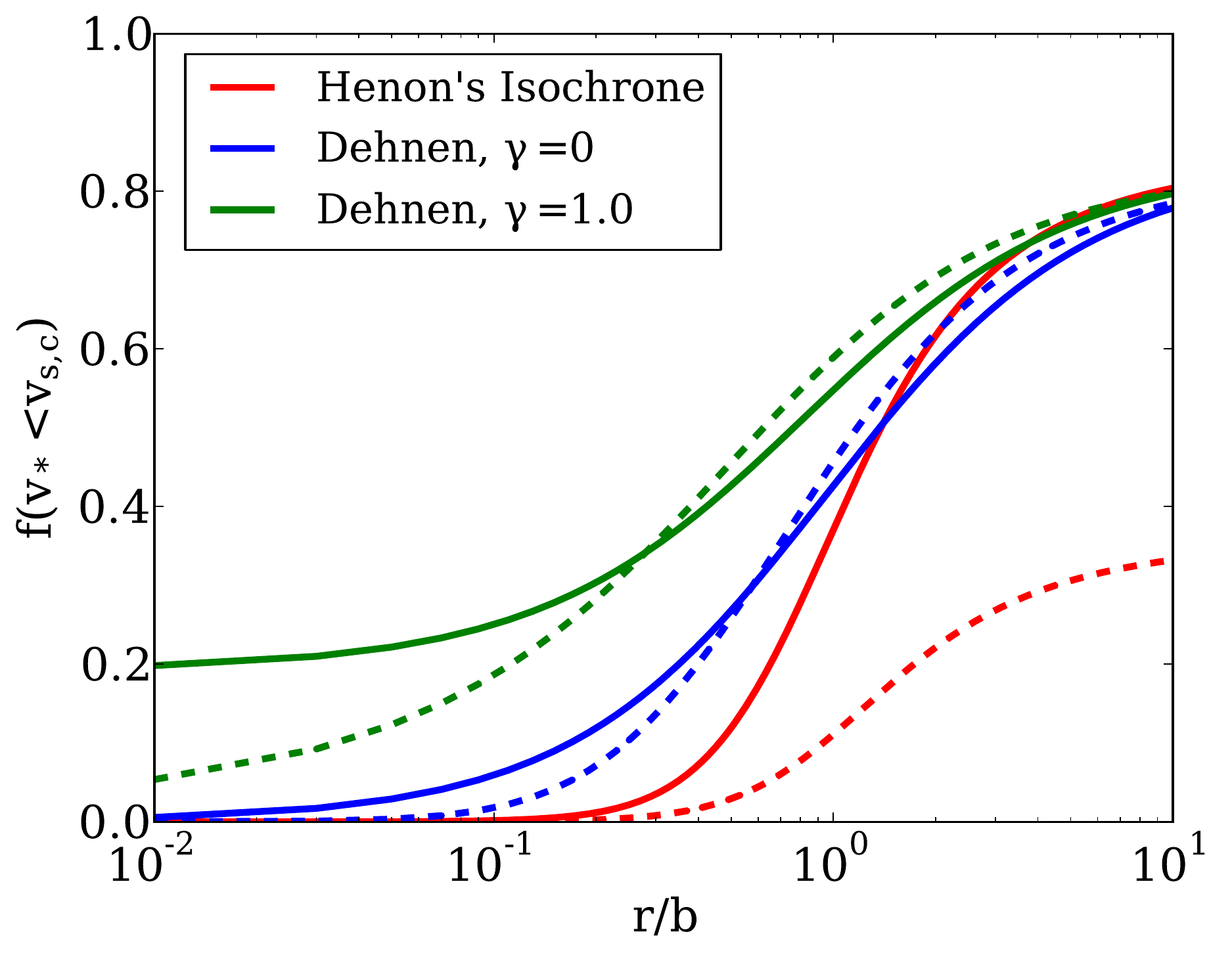}
 \caption{The fraction of stars moving slower than a satellite on a circular orbit, $f(v\sscript{*}<v\sscript{s,c})$, as calculated from the distribution function (solid), and by assuming a Maxwellian distribution of velocities (dashed); as a function of the scale radius for Hénon's Isochrone Model (red) and Dehnen's Model with $\gamma = 0,1.0$ (blue and green respectively), where $v\sscript{s,c}$ is the circular velocity.  As $\sigma$ is larger than the circular velocity inside the entire scale radius of Hénon's Isochrone model, the Maxwellian velocity distribution function severely under-predicts the number of slow moving stars. The radius, $r$ is normalised in units of the scale radius, $b$.} 
 \label{fv.fig}
\end{figure}

To first order, only stars moving slower than the satellite contribute to the friction \citep{Chandrasekhar43} \citep[but see][for an example where the next to leading order term is important]{Antonini12}, and $f(v\sscript{*}<v\sscript{s})$ is taken to be the fraction of stars moving slower than the satellite. Usually a Maxwellian distribution of velocities is assumed, which leads to the simple expression:

\begin{equation}
	f(v\sscript{*}<v\sscript{s}) = \mathrm{erf} (X) - \frac{2X}{\sqrt{\pi}}\exp(-X^2),
	\label{Maxwellain.eqn}
\end{equation}
where $X = v\sscript{S}/\sqrt{2}\sigma$ and $\sigma$ is the velocity dispersion.

This Maxwellian assumption can fail for two reasons. Firstly, it is typically assumed that the velocities of background stars extend to infinity, whereas in realistic backgrounds they will be truncated at the escape velocity, $v\sscript{esc}$. Secondly, the shape of the local velocity distribution function can deviate significantly from the Maxwellian form.

Fig \ref{fv.fig} shows the fraction of stars moving slower than the circular velocity as computed by the Maxwellian approximation and by the distribution function (i.e. the \textit{true} fraction) for Hénon's Isochrone and Dehnen's model with $\gamma=0$ and $\gamma=1.0$. As can be seen, the assumption of a locally Maxwellian velocity distribution works reasonably well (at the $\sim10\%$ level) for both Dehnen models, apart from in the very centre. However, for Hénon's Isochrone model, it gives a very poor match. In particular, the Maxwellian assumption severely under-predicts the fraction of slow moving stars. This, then, is a promising first place to look for understanding why the semi-analytic model in \citet{Petts15}, that assumes a Maxwellian velocity distribution function, fails for a Hénon Isochrone background.

\subsection{Calculating the ``typical interaction velocity'' of background stars, $\bmath{v\sscript{typ}}$}
\label{impact_parameters.ch}

As discussed in section \S \ref{Pett15.ch}, $b\sscript{min}$ depends primarily on the typical interaction velocity, $v\sscript{typ}$. One could postulate that perhaps $v\sscript{typ}$ could be a function of the velocity structure of the background and that $b\sscript{min}$ diverges much faster in a cored profile.

To test this hypothesis we consider that a single background star with velocity $\bmath{v\sscript{*}}$ can minimally and maximally interact with the satellite at velocity:

\begin{subequations}
\begin{align}
&V\sscript{min} = v\sscript{s} - v\sscript{*} \text{ and } V\sscript{max} = v\sscript{s} + v\sscript{*}.\\ 
\intertext{In an isotropic distribution the mean interaction velocity of species $v\sscript{*}$ with the satellite is:}
\bar{V} &= \frac{1}{2}(V\sscript{max}+ V\sscript{min}).
\intertext{It follows that if $v\sscript{*} \le v\sscript{s}$, $\bar{V} = v\sscript{s}$. Integrating over all velocities slower than the satellite:}
v\sscript{typ}&=v\sscript{s} \int_0^{v\sscript{s}}  4 \pi f(v\sscript{*},r)v\sscript{*}^2 {dv\sscript{*}}, \label{vtyp_equation.eq}
\end{align}
\end{subequations}
where $f(v\sscript{*},r)$ is the probability density of a star having velocity $v\sscript{*}$ at $r$. As we are only considering the PDF of the slow stars, the PDF is defined as $\int_0^{v\sscript{s}} 4 \pi f(v\sscript{*})v\sscript{*}^2 {dv\sscript{*}} = 1$. Thus, from equation \ref{vtyp_equation.eq}, $v\sscript{typ} = v\sscript{s}$ as assumed in \citet{Petts15}.

This calculation only holds for isotropic backgrounds, for stars moving slower than the infalling satellite and when one ignores the velocity dependence in $\log(\Lambda)$. These approximations are all used to derive equation \ref{dynfric2.eq}, so $v\sscript{s}$ is the consistent $v\sscript{typ}$ for the standard Chandrasekhar formula. If one wants to include the effects of the stars moving faster than the satellite, $v\sscript{typ} \ne v\sscript{s}$ and equation \ref{dynfric2.eq} is inadequate. In this case one must integrate a more general form of Chandrasekhar's formula. We discuss this in section \S \ref{fast_stars.ch}.

\subsection{Correctly capturing the stalling effect in large cores}
\label{core_stalling.ch}

In \citet{Petts15}, we argued that core stalling occurs when $b\sscript{min} \geq b\sscript{max}$ and/or the fraction of slow moving stars at the satellite's position approaches zero as it inspirals.  Indeed, this gave an excellent match to core-stalling in a Dehnen background, even for $\gamma = 0$. However, for large and particularly flat cores, the \citet{Petts15} model fails. This owes in part  to the poor approximation of a Maxwellian velocity distribution function for the Hénon Isochrone background. However, as we shall show in section \S \ref{results.ch}, this is only part of the story.

To obtain the correct stalling radii for large cores, we extend an idea already presented in \citet{Goerdt10} that core-stalling occurs when the infalling satellite tidally disrupts the cusp and forms its own small core. The authors claim that the stalling occurs after core creation due to the mechanism described in R06, whereby the stars move in epicycles in the combined potential of the galaxy and satellite without a net change in energy when averaged over the orbit. They state that this transformation must occur at approximately the tidal radius, where the satellite itself tidally strips the cusp of the galaxy. The authors found the empirical relation:

\begin{equation}
	r\sscript{s}\sim(2-\gamma) r\sscript{t},
	\label{Goerdt10_relation.eq}
\end{equation}
where $r\sscript{s}$ is the stalling radius of satellite, and $r\sscript{t}$ is the tidal radius of the satellite. The coefficient was derived empirically for inner slopes of $\gamma = 0.5, 1.75$. However, we show that although this coefficient gave an excellent fit, it is an artifact arising from an inaccurate definition of the tidal radius. The formal definition of the tidal radius for a point mass on a circular orbit is \citep{King62,BT08}:

\begin{equation}
	r\sscript{t}^3 = \frac{GM\sscript{s}}{\Omega^2 - \frac{\rm{d}^2\Phi}{\rm{d}r^2}},
	\label{rt.eq}
\end{equation}
where $\rm{d}^2\Phi/\rm{d}r^2$ is the second derivative of the host galaxy's potential at the satellite's position, and $\Omega$ is the rotational velocity of the satellite. For a circular orbit, by definition:

\begin{equation}
	\Omega^2 = \frac{GM\sscript{enc}}{r^3}.
	\label{omega.eq}
\end{equation}

Equation \ref{rt.eq} highlights why the tidal radius becomes very large near the centre of galaxies with a large core, as both terms in the denominator tend to zero as $r \rightarrow 0$. Conversely, for very cuspy distributions the mass is very centrally concentrated and thus the denominator greatly increases towards the centre of the system.

Inside a homogeneous spherical galaxy $\Omega^2=\rm{d}^2\Phi/\rm{d}r^2=(4/3)\pi G \rho\sscript{0}$ and $r\sscript{t} = \infty$ everywhere, independent of the satellite orbit. With an infinite tidal radius, any star in the galaxy is formally bound to the satellite and transfers no net energy when time averaged over its orbit, thus the satellite experiences no friction (consistent with the similar analytic argument of R06). However, such a configuration would be unstable due to the influence of the satellite on the background. If the assumptions of homogeneity are relaxed and we consider a realistic and finite density profile with a large core, such as Hénon's Isochrone model, the tidal radius is now only zero at the very centre of the system and finite everywhere else. However, the presence of the large core can cause the tidal radius to grow very large as the cluster migrates into the core and $\alpha \rightarrow 0$, causing the cusp/core transformation described in \citet{Goerdt10} to occur when $M\sscript{s} \gg M\sscript{enc}$.

The dynamics of stars within the tidal radius are dominated by the satellite as opposed to the background and the phase-space distribution of the background will be drastically disrupted from its original state when $r = r\sscript{t}$.

At this scale the galactic centre is tidally disrupted by the satellite, reshaping the velocity distribution of the core and stalling the orbit.  The probability of stars scattering off of the satellite at a specific angle is no longer uniform and $v\sscript{typ} \neq v\sscript{s}$, as some relative interaction velocities become more probable than others. Thus, the assumption of an isotropic pristine core is broken and equation \ref{dynfric2.eq} fails. Calculating $v\sscript{typ}$ of the resulting distribution is far from trivial, especially because the combined potential is not spherical and is only stationary in the co-rotating frame of the satellite.

R06 explain the stalling behaviour by showing that for a harmonic core there exist states where no net energy is transferred to the satellite. \citet{Inoue11} showed that in this regime a few particles have orbits that feed energy to the satellite over a few satellite orbits. These so called ``horn'' particles have orbits that stay close to the tidal radius of the satellite for extended periods \citep[namely between the L2 and L3 equipotential surfaces, see figure A1 of][]{Inoue11}. These horn particles counter-act most of the dynamical friction due to a complex 3-body interaction with the satellite and the galaxy. The horn particles evidently play a vital role in keeping the satellite buoyant. However, particles occupying this region of phase space do eventually move away from the satellite and other particles enter the horn trajectory \citep[see fig. 10 and table 1 of][]{Inoue11}.  We consider that the analytic estimate of R06 may be degenerate with the presence of horn particles if most stars that come close to the tidal radius of the satellite in this new distribution will be horn trajectory for at least part of their orbits.

As an ansatz, we put a constraint on the frictional force so that:
\begin{equation}
	\dot{\bmath{v}}\sscript{df} =
	\begin{cases}
            \frac{d\bmath{v\sscript{S}}}{dt}, &         \text{if } r\sscript{t}(r\sscript{a}) < r\sscript{a}\\
            0, &         \text{if } r\sscript{t}(r\sscript{a}) \geq r\sscript{a},
    \end{cases}
    \label{tidal_stalling.eq}
\end{equation}
where $\frac{d\bmath{v\sscript{S}}}{dt}$ is the frictional model employed (either equation \ref{dynfric2.eq}, or \ref{dynf_full.eq}). Hereafter, we call this mechanism ``tidal stalling''. In section \S \ref{results.ch}, we will show that with this additional constraint core stalling can be captured remarkably well in both large cores and cuspy galaxies. This suggests that the stalling in large cores occurs via the same ``cusp disruption'' mechanism that occurs in cuspy profiles, in agreement with \citet{Goerdt10}. The only ``unique'' aspect of a large pre-existing core is that the extended tidal influence of the satellite in the shallow region means that the satellite can disrupt the galactic centre at $M\sscript{enc} \gg M\sscript{S}$.

\subsection{The effect of fast moving stars}
\label{fast_stars.ch}

When deriving equation \ref{dynfric.eq}, \citet{Chandrasekhar43} assumed that only stars moving slower than the satellite contribute to the frictional force. In most distributions this is a good approximation as there is an abundance of slow moving stars that all contribute to, and dominate, the friction. The effect of interactions with faster moving stars is fundamentally different, which we demonstrate by considering the general Chandrasekhar formula (equations. 25 and 26 in \citet{Chandrasekhar43}):

\begin{align}
	\frac{d\bmath{v\sscript{S}}}{dt} &= - 4 \pi G^2 M\sscript{s} \rho(r) \frac{\mathbf{v\sscript{s}}}{v\sscript{s}^3} \int_0^{\sqrt{-2\phi(r)}} \frac{J(v\sscript{*})}{8v\sscript{*}}  4 \pi f(v\sscript{*})  v\sscript{*}^2 {dv\sscript{*}}, \label{dynf_full.eq}\\
	J(v\sscript{*}) &= \int_{|v\sscript{s}-v\sscript{*}|}^{v\sscript{s}+v\sscript{*}}  \left( 1 + \frac{v\sscript{s}^2 - v\sscript{*}^2}{V^2}\right) \log\left(1 + \frac{b\sscript{max}^2 V^4}{G^2 M\sscript{s}^2}\right) {dV},\label{J.eq}
\end{align}
where $V$ is the relative velocity of the encounter, and $J(v\sscript{*})$ is a function describing the interaction strength of a single velocity species integrated over all possible relative velocities and impact parameters. $J(v\sscript{*})$ is positive for all $V$ if $v\sscript{*} < v\sscript{s}$, therefore all slow moving stars in the system remove energy from the satellite.

Intriguingly equation \ref{J.eq} predicts ``dynamical buoyancy'' from a portion of the stars moving faster than the satellite. If $v\sscript{*} > v\sscript{s}$ then $J(v\sscript{*})$ is negative if:

\begin{equation}
	\frac{v\sscript{s}^2 - v\sscript{*}^2}{V^2} < -1.
	\label{buoyant_criteria.eq}
\end{equation}
Such interactions feed energy into the satellite producing a buoyancy effect opposing the frictional force of other fast moving stars. However, as the minimum impact parameter is smaller for interactions with a higher relative velocity, when summed over all impact parameters there is a net residual frictional force from the fast moving stars. This residual force is usually small compared to the friction coming from the slow moving stars, and is ignored in deriving equation \ref{dynfric.eq}. However, \citet{Antonini12} showed that in situations where the density of fast moving stars is much greater than that of the slow moving stars this residual force can become non-negligible or even dominate in extreme conditions (such as deep in the potential well a of shallow stellar cusp around a super massive black hole). Subsequently, \citet{ArcaSedda14} modelled the dynamical friction on satellites in galaxies of various inner log-slope, $\gamma$, taking into account the non-locality of the friction in cusps as well as the contribution of the fast stars, showing improved agreement in galactic centres compared with using equation \ref{dynfric.eq}. Similarly, inside a large core there are very few stars moving slower than the circular velocity, and the residual friction from the fast stars could be important in this case.

In section \S \ref{results.ch} we solve equation \ref{dynf_full.eq} in addition to equation \ref{dynfric2.eq} to quantify the effect of these fast moving stars. The possible role of the fast moving stars in the stalling phase is discussed in section \ref{fast_stalling.ch}.

\subsection{Summary of the updated model}

In Table \ref{dynf_models.tbl} we briefly summarise the differences in our updated models as compared to our model presented in \citet{Petts15}. The new models use $b\sscript{max}$ and $b\sscript{min}$ as described in P15, with the exception of the P16f model which includes the relative velocity dependence of interactions in $\log(\Lambda)$. In general, the P16f model will give the most physically accurate results, however it requires a double integral as opposed to the single integral P16 requires. For most distributions P16 is adequate, however the fast moving stars can make up a significant portion of the friction in regions where there are few slow moving stars available, as we will show in section \S \ref{results.ch}. We notate the optional inclusion of tidal stalling in the models with the addition of ``+TS'' to the name (i.e. with tidal stalling switch on P16 and P16f become P16+TS and P16f+TS, respectively).

\begin{table}
\begin{minipage}{85mm}

  \begin{tabular}{@{}llll@{}}
  \hline
    model & P15& P16 & P16f\\
    \hline
    Equation & \ref{dynfric.eq} & \ref{dynfric2.eq} & \ref{dynf_full.eq} \\
    $b\sscript{max}$ & $\rm{min}\left(\frac{\rho(r)}{\nabla\rho},r\right)$&$\rm{min}\left(\frac{\rho(r)}{\nabla\rho},r\right)$&$\rm{min}\left(\frac{\rho(r)}{\nabla\rho},r\right)$ \\
    $b\sscript{min}$&$\rm{max}(\frac{GM\sscript{s}}{v\sscript{s}^2}, r\sscript{hm})$&$\rm{max}(\frac{GM\sscript{s}}{v\sscript{s}^2}, r\sscript{hm})$&$\rm{max}(\frac{GM\sscript{s}}{V^2}, r\sscript{hm})$ \\
    $f(v\sscript{*})$ & Maxwellian & Self-Consistent & Self-Consistent\\
    Stalling & $b\sscript{max} = b\sscript{min}$& (+Tidal Stalling) & (+Tidal Stalling)\\
    Stars&$v\sscript{*}<v\sscript{s}$&$v\sscript{*}<v\sscript{s}$&$v\sscript{*}<v\sscript{esc}$\\
    \hline
\end{tabular}
\caption{Parameters of three different dynamical friction models. From left to right, these are: the P15 model; the P15 model with with correct background distribution function (P16+TS with ``tidal stalling'', P16 without); and the P16 model including the velocity dependence of $\log(\Lambda)$ and the effect of stars that move faster than the satellite (P16f+TS with ``tidal stalling'', P16f without). A ``Self-Consistent'' $f(v\sscript{*})$ means using the isotropic distribution function as calculated from the Eddington formula \citep[e.g.][]{BT08}, rather than the more usual Maxwellian approximation. The P16 and P16f models will stall if $b\sscript{max} = b\sscript{min}$, as in the P15 model, but introduce an additional ``tidal stalling'' mechanism, as discussed in \S 3.5.}
\label{dynf_models.tbl}

\end{minipage}
\end{table}

\section{Simulations}
\label{simulations.ch}

In order to test our predictions, we use the tree-code GADGET2 \citep{Springel05} to simulate the inspiral of satellites in Hénon's Isochrone model and Dehnen's model. We use units of $G=M\sscript{g}=b=1$ for Hénon's model, where $G$ is the gravitational constant, $M\sscript{g}$ is the total galaxy mass and $b$ is the scale radius of the galaxy given. We use point mass satellites with masses of multiples of $1.595\times10^{-4} M\sscript{g}$, which corresponds to $2\times10^5 \unit{M\sscript{\odot}}$ in the Hénon model when normalised to the same central density as the simulations in R06. If the stalling mechanism is independent of the velocity structure we should obtain similar ratios of $M\sscript{s}/M\sscript{enc}$ at the stalling radius to R06. The initial conditions of the simulations are displayed in Table \ref{IC.tbl}.

We compare GADGET2 simulations to our semi-analytical model where we use a static analytic model for the background galaxy and perturb the orbit with different friction models as described in Table \ref{dynf_models.tbl}. We use a leap-frog integrator with variable time-step to integrate the perturbed orbit, which conserves energy to a relative error of $< 10^{-7}$ in the absence of dynamical friction over the timescales considered. With dynamical friction switched on, if we sum up the removed orbital energy and add it to the final energy of the satellite we obtain the same relative error.

We use the following naming convention for simulations: for $N$-body models computed with GADGET2 we name the simulation ${\tt gt\_\left<IC\right>}$, where ${\tt \left<IC\right>}$ are the initial conditions described in table \ref{IC.tbl}. For semi-analytic models we name the simulation ${\tt df\_\left<X\right>\_\left<IC\right>}$, where ${\tt \left<X\right>}$ is the dynamical friction model used, as described in table \ref{dynf_models.tbl}.

\begin{table}
\begin{minipage}{85mm}

  \begin{tabular}{@{}lrrrrrl@{}}
  \hline
   IC Name  & $M\sscript{s}$ & $r\sscript{0}$& $v\sscript{0}$ & $\gamma$& $N\sscript{bg}$\\
   & (1.595e-4$\unit{M\sscript{g}}$) & ($b$)& ($v\sscript{c}$) &&\\
 \hline
   H & 1 & 1&1.0& -&$2^{24}$\\
   H2 & 2 & 1&1.0& -&$2^{24}$\\
   H2e4 & 2 &1 &0.4&- &$2^{24}$\\
   H2multi* & 2 & 0.5&1.0& -&$2^{24}$\\
   H4 & 4 & 1&1.0& -&$2^{24}$\\
   H1.5b & 1 & 1.5&1.0&- &$2^{24}$\\
   H0.3b & 1 & 0.295&1.0&- &$2^{24}$\\
   H0.17b & 1 & 0.168&1.0&- &$2^{24}$\\
   D0 & 2 & 0.5 & 1.0& 0& $2^{22}$\\
   D1 & 2 & 1 & 1.0& 1&$2^{22}$\\
  \hline

\end{tabular}
\caption{Initial conditions of the simulations. Column 1 lists the name of the initial conditions. Column 2 states the mass of the satellite in units of $1.595\times10^{-4} M\sscript{g}$. Column 3 shows the initial position of the satellite, where $r\sscript{0}$ is in units of the scale length, $b$, in equations \ref{Henon_rho.eq} and \ref{Dehnen_rho.eq}. Column 4 shows the initial satellite velocity, $v\sscript{0}$, in units of $v\sscript{c}$, the circular velocity at $r\sscript{0}$. Column 5 shows the inner asymptotic slope of the Dehnen models and column 6 shows the number of particles used to simulate the halo in the GADGET2 simulations. (*H2multi includes two satellites, initially orbiting in the x-y and x-z planes.)}
\label{IC.tbl}

\end{minipage}
\end{table}

\section{Results}
\label{results.ch}

\subsection{Circular inspiral}

\begin{figure}
 \includegraphics[width=\linewidth]{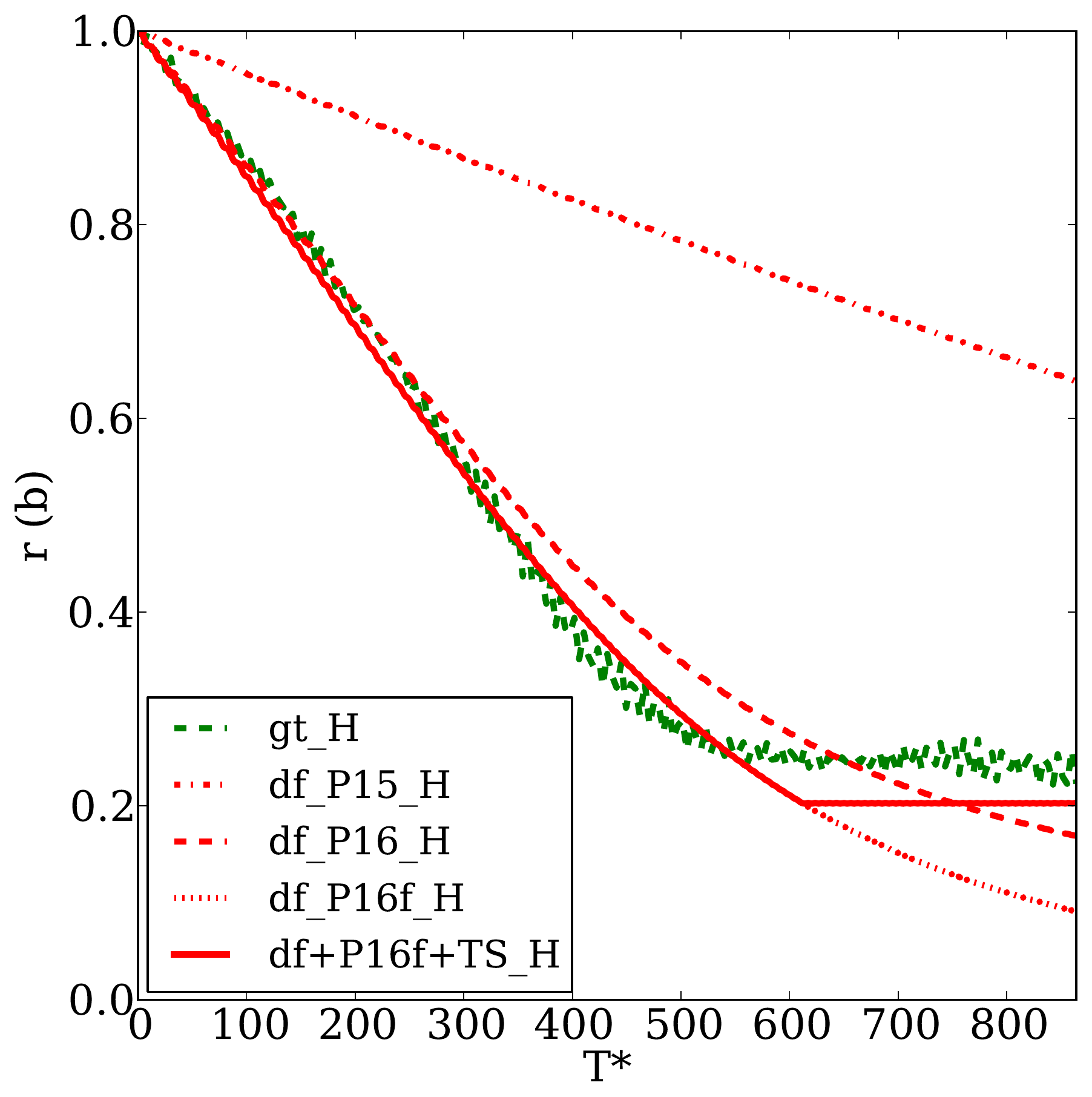}
 \caption{Evolution of the satellite orbit with time for simulations ${\tt gt\_H}$ (green line) and ${\tt df\_\left<X\right>\_H}$. The ${\tt df\_\left<X\right>\_H}$ simulation is shown with the standard Maxwellian approximation (P15; dot-dashed red line), with the true $f(v\sscript{*}<v\sscript{s})$ (P16; dashed red line), with the true $f(v\sscript{*}<v\sscript{s})$ including the effects of the fast moving stars (P16f; dotted red line), and the same model with tidal stalling (see section \S \ref{core_stalling.ch}) turned on (P16f+TS; solid red line) }
 \label{2e5_fv_max.fig}
\end{figure}

\begin{figure}
 \includegraphics[width=\linewidth]{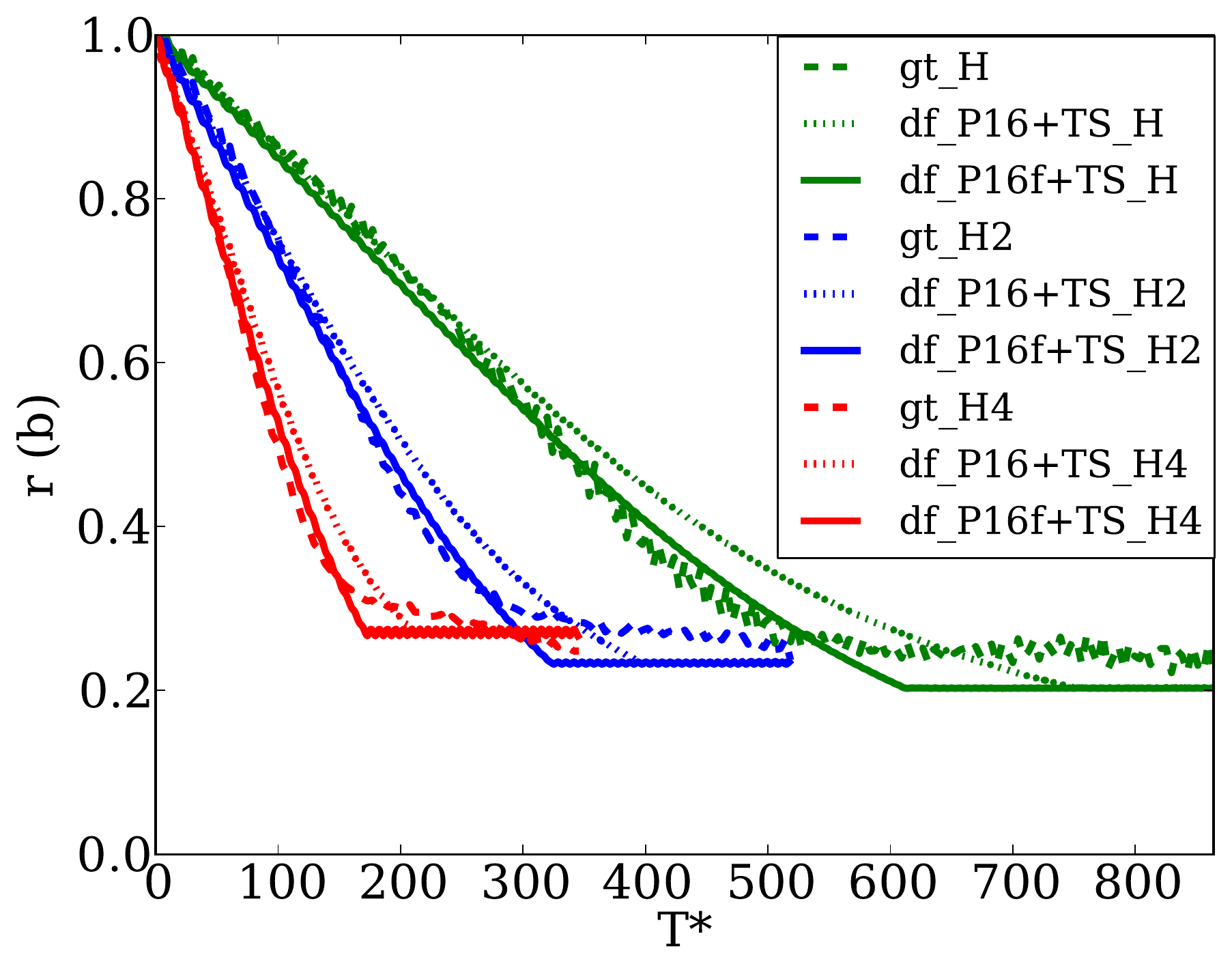}
 \caption{Evolution of the satellite orbit with time for simulations computed with GADGET2 (${\tt gt\_H}$, ${\tt gt\_H2}$, ${\tt gt\_H4}$; dashed lines) and our model considering only the slow moving stars (P16+TS; dotted lines) and considering all the stars (P16f+TS; solid lines). The satellite masses are $1.595\times10^{-4}$, $3.19\times10^{-4}$ and $6.38\times10^{-4}$ for simulations ${\tt gt\_H}$, ${\tt gt\_H2}$ and ${\tt gt\_H4}$, respectively. The simulation initial conditions are described in table \ref{IC.tbl}.} 
 \label{mass_circ.fig}
\end{figure}

Fig. \ref{2e5_fv_max.fig} shows $N$-body simulation ${\tt gt\_H}$ and 4 realisations of the semi-analytic model, ${\tt df\_\left<X\right>\_H}$, with different force models. The dot-dashed red line shows the result obtained by the standard Maxwellian approximation (P15), and gives an extremely poor fit to the $N-$body data. When the true $f(v\sscript*)$ for Hénon's isochrone is used (P16; dashed red line) then the inspiral is reproduced excellently for the majority of the orbital evolution, as the velocity distribution has the correct shape and the fraction of slow moving stars is no longer under-predicted. Deep in the core, prior to stalling, equation \ref{dynfric2.eq} slightly underestimates the friction experienced by the satellite in the $N-$body model. Solving equation \ref{dynf_full.eq} (P16f; dotted red line) shows that the discrepancy originates from ignoring the residual friction from the fast moving stars, which becomes significant in this region. Including the fast moving stars gives an excellent fit right down to the stalling radius.

In our semi-analytic models P16+TS and P16f+TS, dynamical friction stops when $r\sscript{t}(r\sscript{a}) = r\sscript{a}$, when the satellite can tidally disrupt the core. At this point the satellite stalls, with inspiral being much slower than one would estimate if the core is assumed to resemble its initial conditions (marked with a red dashed line and a red dotted line for equations \ref{dynfric2.eq} and \ref{dynf_full.eq} respectively). This simple model for the tidal stalling gives a very good fit to the $N-$body data, which stalls at $M\sscript{s}/M\sscript{enc} = 0.03$. This is a factor of 2 smaller than in R06, which is what we expect as Hénon's Isochrone has a shallower core than the model R06 use. The semi-analytic model stalls at  $M\sscript{s}/M\sscript{enc} = 0.04$, slightly farther in. It is not surprising that we underestimate the tidal radius with equation \ref{rt.eq}, as it is derived under the assumption that $r\sscript{t} \ll r$, which allows one to linearise the forces. Nevertheless, the approximate tidal radius gives a satisfying fit.

Fig. \ref{mass_circ.fig} shows how the $N$-body simulations and semi-analytic models vary as a function of the satellite mass. The semi-analytic models reproduce the inspiral excellently. The stalling radii scale in the same way as the $N$-body models, with larger masses stalling further out. Although the tidal radii scale as $M\sscript{S}^{1/3}$, the stalling radii have a sub $M\sscript{S}^{1/3}$ scaling, as the other terms in equation \ref{rt.eq} also depend on $r$.

One could fit a free parameter that scales with the mass to better reproduce the stalling radii, however, any such free parameter would be dependant on the galaxy model. We choose to keep the model free of any free parameters to ensure its predictive power in general potentials.

\subsubsection{Effect of initial distance}

\label{initial_distance.ch}

Simulations ${\tt gt\_H1.5b}$, ${\tt gt\_H0.3b}$ and ${\tt gt\_H0.17b}$ have the same initial conditions at ${\tt gt\_H}$, except the satellites are initially on circular orbits at $1.5 \unit{b}$, $0.295\unit{b}$ and $0.168 \unit{b}$, respectively.

\begin{figure}
 \includegraphics[width=\linewidth]{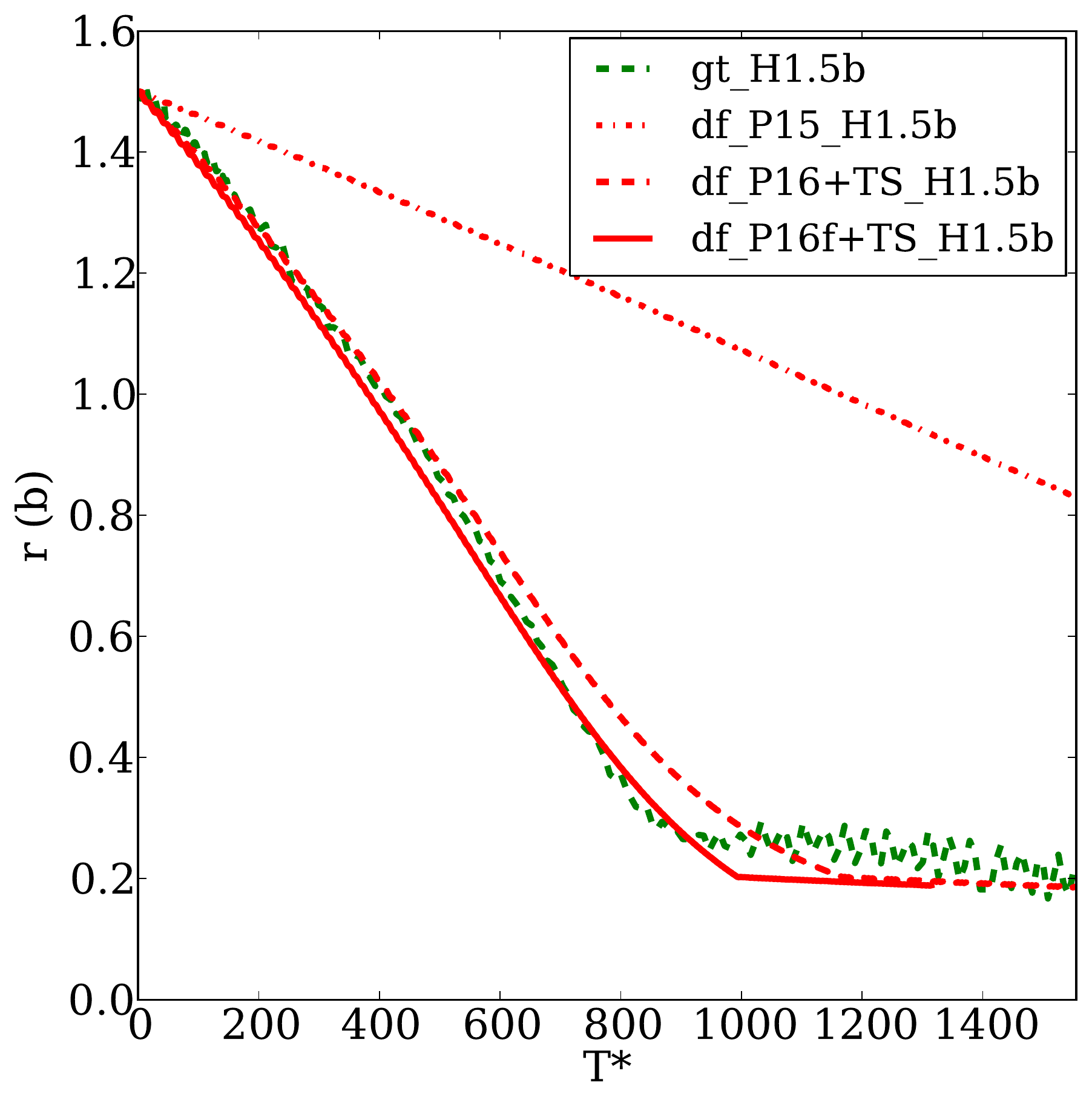}
 \caption{Evolution of the satellite orbit with time for simulations ${\tt gt\_H1.5b}$ (green line) and ${\tt df\_\left<X\right>\_H1.5b}$. The ${\tt df\_\left<X\right>\_H1.5b}$ simulation is shown with the standard Maxwellian approximation (P15; dot-dashed red line), with the true $f(v\sscript{*}<v\sscript{s})$ (P16+TS; dashed red line), and with the true $f(v\sscript{*}<v\sscript{s})$ including the effects of all stars (P16f+TS; solid red line). The later two models include the effects of tidal stalling.}
 \label{Hout.fig}
\end{figure}

In ${\tt gt\_H1.5b}$ the satellite is initially far outside the core, where the local density slope, $\gamma = -2.4$. Fig. \ref{Hout.fig} shows that in models using the self-consistent $f(v\sscript{*})$, the inspiral is very well reproduced throughout the satellite's migration to the cored region. As well as using the correct distribution function, the success of the model owes also to our radially varying $b\sscript{max}$, which is smaller than $r$ in the cuspy outer regions \citep[see][for more detail]{Just11,Petts15}. The satellite stalls at the same radius as in ${\tt gt\_H}$, verifying that the stalling radius is independent of the initial distance if the satellite originates from outside the core region, in agreement with \citet{Goerdt10}.

Simulation ${\tt gt\_H0.3b}$ starts just outside of where the satellite stalls in simulations ${\tt gt\_H}$ and ${\tt gt\_H1.5b}$, but stalls slightly further in. This is because the distribution function is in its pristine state and the satellite feels friction until it has enough time for the background and satellite to settle into the R06 state with no net momentum exchange. The same is true for ${\tt gt\_H0.17b}$, where the satellite is initially below the stalling radius of the satellites in ${\tt gt\_H}$ and ${\tt gt\_H1.5b}$. In both ${\tt gt\_H0.3b}$ and ${\tt gt\_H0.17b}$ the semi-analytic model including all stars is initially in great agreement with the $N$-body results, as the distribution function assumed in the model is initially correct. This strengthens the point that a shift in the distribution function is why equations \ref{dynfric2.eq} and \ref{dynf_full.eq} fail without our additional stalling prescription. Interestingly, we do not see the ``dynamical buoyancy'' effect discovered in \citet{Cole12} in simulation ${\tt gt\_H0.17b}$, this is discussed in section \S \ref{buoyancy.ch}.

One should note that although the semi-analytic formula will be poor at reproducing $N$-body results if $r\sscript{i} \sim r\sscript{t}(r\sscript{i})$, where $r\sscript{i}$ is the initial apocentre, this is a purely numerical effect. In the real universe, initial conditions such as ${\tt gt\_H0.3b}$ and ${\tt gt\_H0.17b}$ are impossible as the galaxy potential will be self-consistent with the presence of the satellite upon its formation, meaning that if the satellite is born deep in the cored region it will initially be in a stalled state. As such, our semi-analytic model for ${\tt gt\_H0.17b}$ in which the satellite simply has no orbital evolution, is justified.

We would also like to note that in theory one may have no need to employ equation  \ref{tidal_stalling.eq} if one could include the self consistent velocity distribution that includes the effect of the satellite on the distribution function of the background. However such a model is far from trivial to calculate since the satellite is off centre from the background distribution, thus the distribution function is highly aspherical around the satellite. Such a model is beyond the scope of this work. We use our tidal stalling prescription as a simple physically motivated model to capture the stalling radius at which the distribution function should be heavily perturbed by the satellite.

\begin{figure}
 \includegraphics[width=\linewidth]{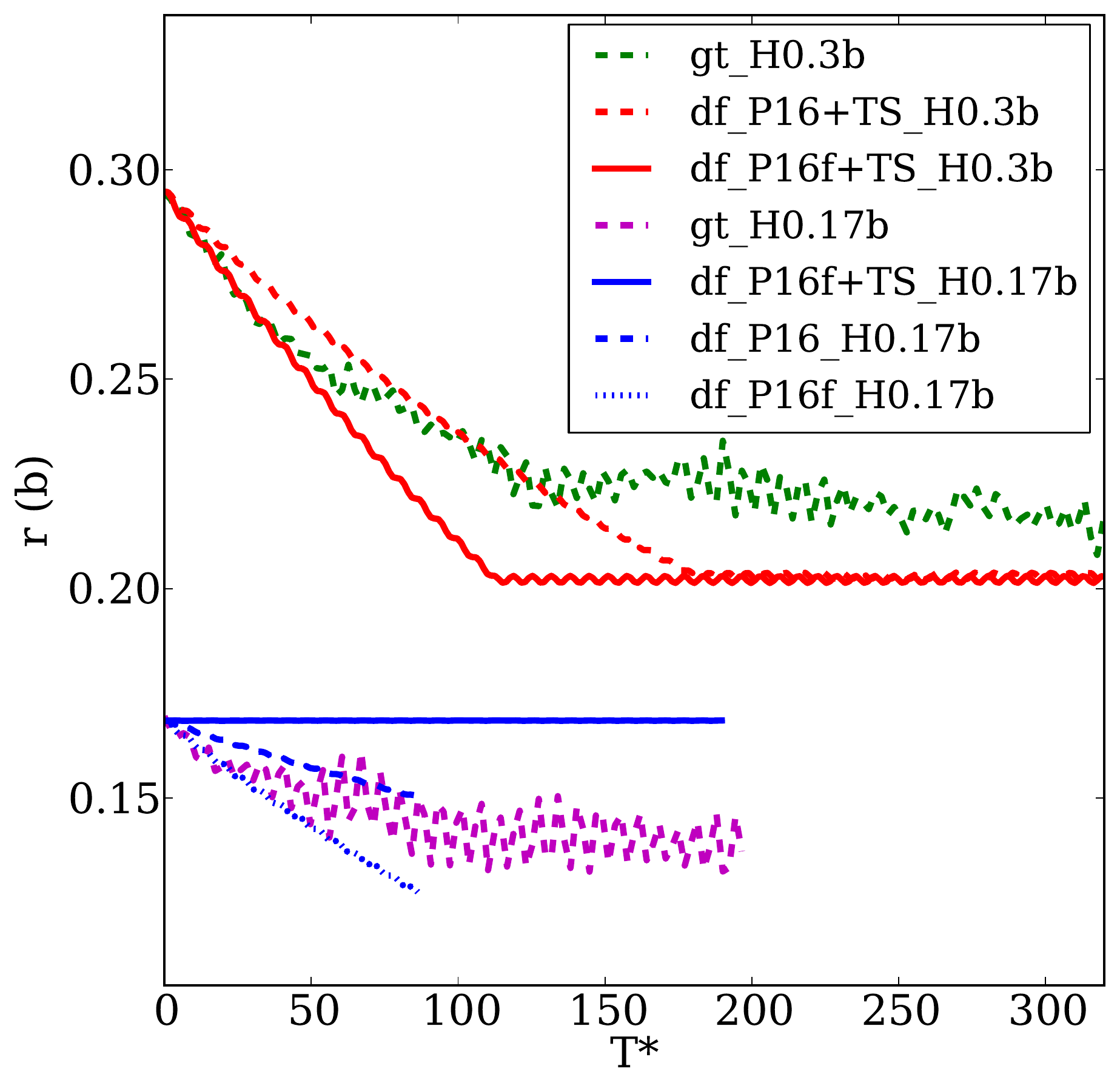}
 \caption{Evolution of the satellite orbit with time for simulations ${\tt gt\_H0.3b}$ (green dashed line), ${\tt df\_\left<X\right>\_H0.3b}$ (red lines), ${\tt gt\_H0.17b}$ (magenta dashed line) and ${\tt df\_\left<X\right>\_H0.17b}$ (blue lines). Simulations ${\tt df\_\left<X\right>\_H0.3b}$ and ${\tt df\_\left<X\right>\_H0.17b}$ are realised with different dynamical friction models, as specified in the legend.}
 \label{Hin.fig}
\end{figure}

\subsection{Elliptical inspiral}

We re-ran simulations ${\tt gt\_H2}$ and ${\tt df\_H2}$ with an initial tangential velocity of $0.4 \unit{v\sscript{c}}$ (simulations ${\tt gt\_H2e4}$ and ${\tt df\_H2e4}$).

Taking the stalling radius to be the same as the radius at which a circular orbit at apocentre would stall gives good agreement to the $N$-body model. This makes intuitive sense if one considers that stalling is a result of tidal disruption of the core. Prior to stalling, the satellite experiences friction when passing through pericentre, as the satellite moves quickly in and out of the core. However, once the entire orbit is inside the core the satellite can tidally disrupt the core over the course of a few orbits. For a spherical potential the apocentre can easily be calculated from any point of the orbit by solving the equation for the turning points of the orbit \citep{BT08}:

\begin{equation}
	r^{-2} + \frac{2.0\left[\Phi(r) - E\right]}{L^2} = 0,
\end{equation}
where $\Phi$, $E$ and $L$ are the potential, specific orbital energy and specific angular momentum, respectively. The largest and smallest solutions are the apocentre and pericentre of the orbit, respectively.

\begin{figure}
 \includegraphics[width=\linewidth]{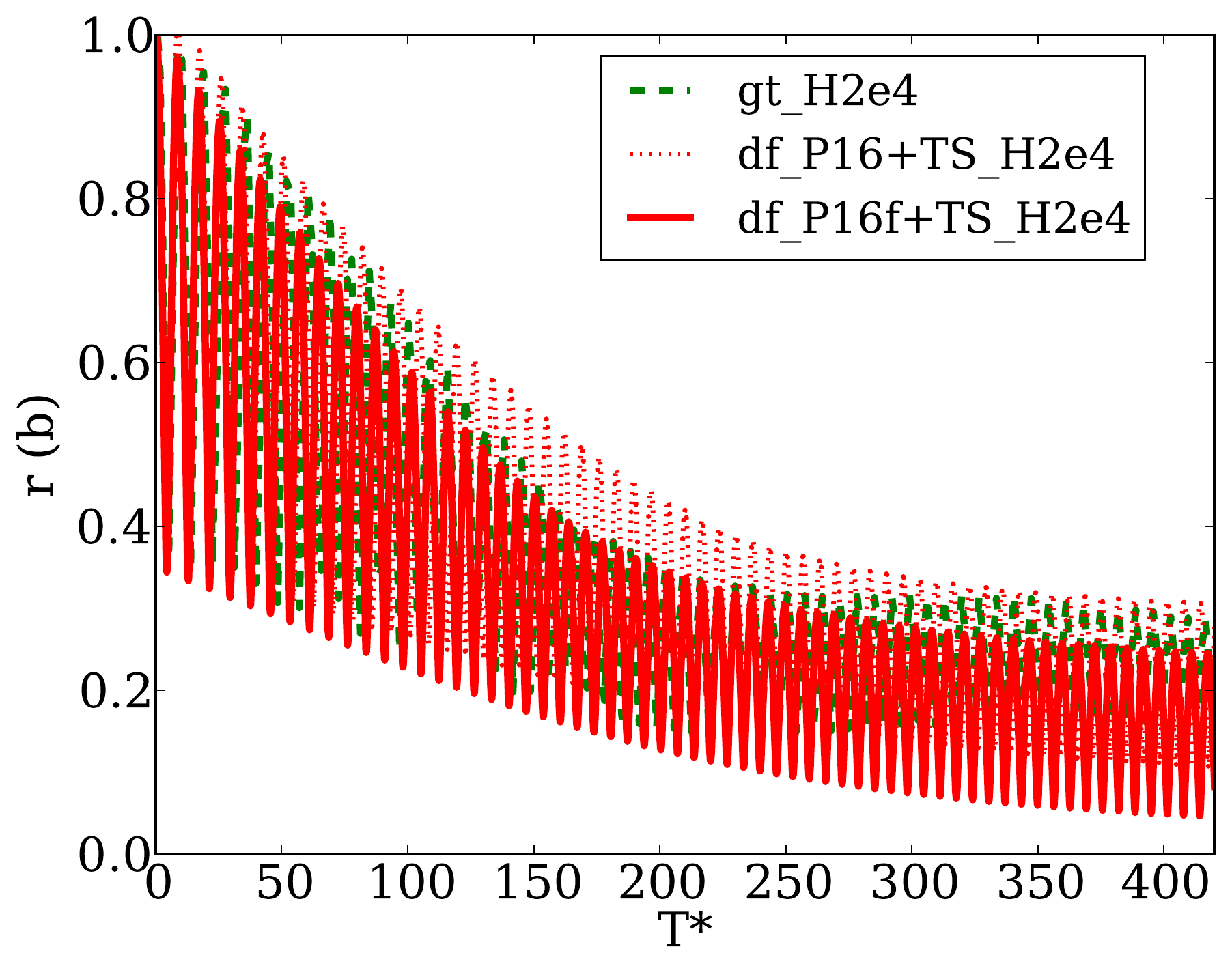}
 \caption{Evolution of the satellite orbit with time for simulations ${\tt gt\_H2e4}$ (green line), ${\tt df\_P16+TS\_H2e4}$ (red dotted line) and ${\tt df\_P16+TS\_H2e4}$ (solid red line).}
 \label{ecc.fig}
\end{figure}

\subsection{Updated friction model in weak and strong cusps}

\begin{figure}
 \includegraphics[width=\linewidth]{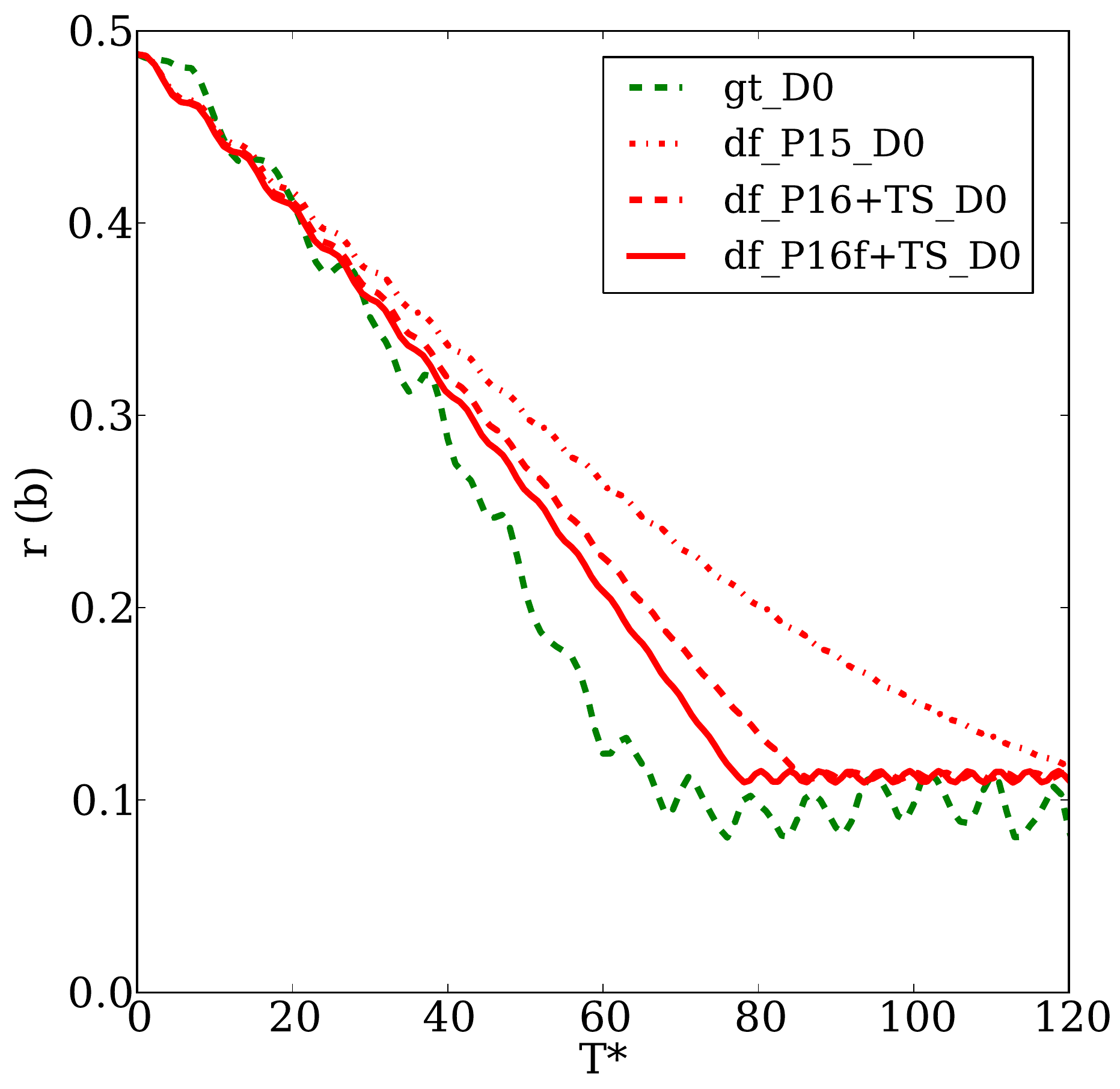}
 \caption{Evolution of the satellite orbit with time for simulations ${\tt gt\_D0}$ (green line) and ${\tt df\_\left<X\right>\_D0}$. The ${\tt df\_\left<X\right>\_D0}$ simulation is shown with the standard Maxwellian approximation (P15; dot-dashed red line), with the true $f(v\sscript{*}<v\sscript{s})$ (P16+TS; dashed red line), and with the true $f(v\sscript{*}<v\sscript{s})$ including the effects of all stars (P16f+TS; solid red line). The later two models include the effects of tidal stalling.}
 \label{g0.fig}
\end{figure}

\begin{figure}
 \includegraphics[width=\linewidth]{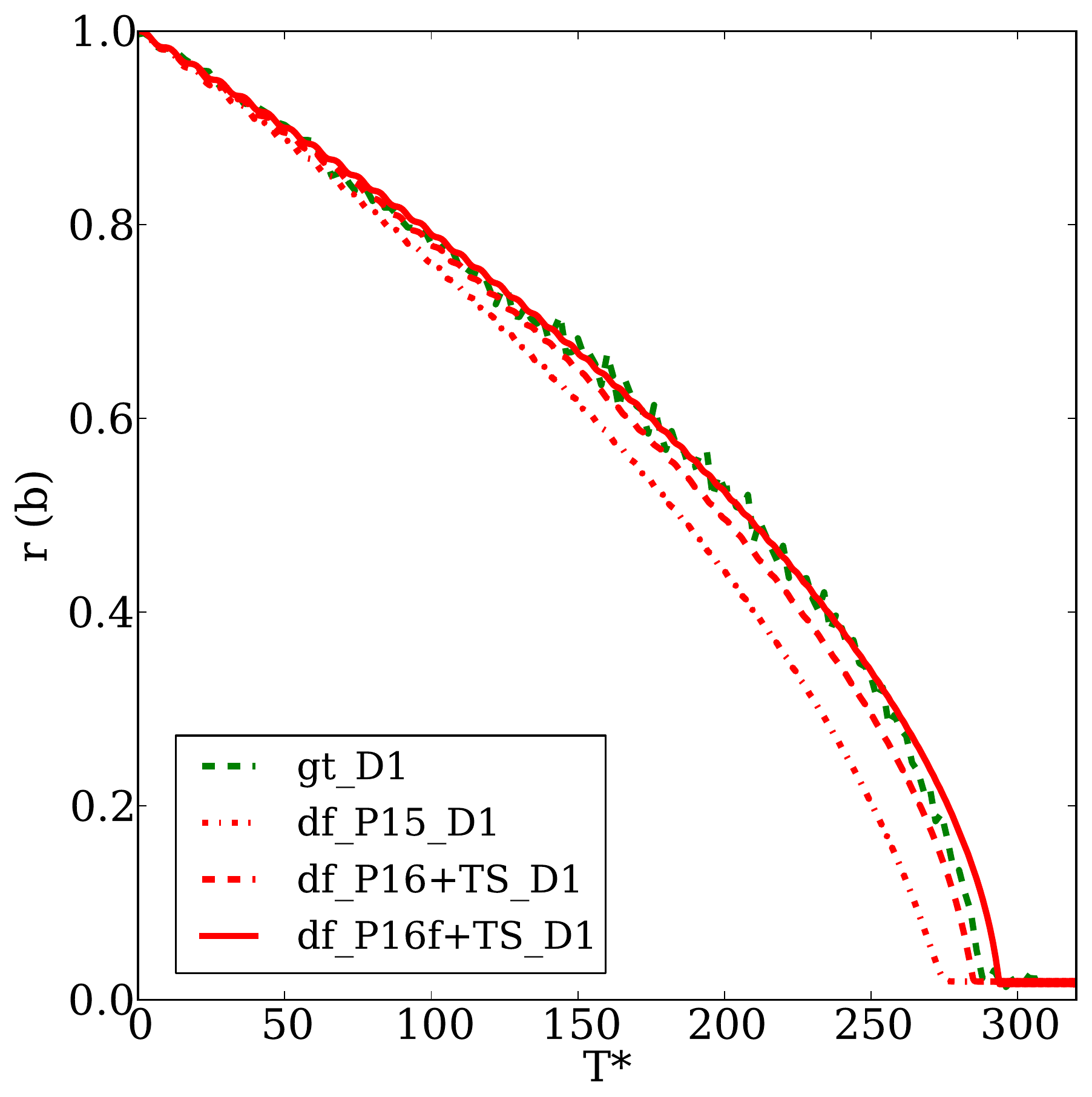}
 \caption{Evolution of the satellite orbit with time for simulations ${\tt gt\_D1}$ (green line) and ${\tt df\_\left<X\right>\_D1}$. The ${\tt df\_\left<X\right>\_D1}$ simulation is shown with the standard Maxwellian approximation (P15; dot-dashed red line), with the true $f(v\sscript{*}<v\sscript{s})$ (P16+TS; dashed red line), and with the true $f(v\sscript{*}<v\sscript{s})$ including the effects of all stars (P16f+TS; solid red line). The later two models include the effects of tidal stalling.}
 \label{g1.fig}
\end{figure}

In \citet{Petts15} our dynamical friction model used the Maxwellian approximation and satellites stalled when $b\sscript{min} \le b\sscript{max}$. As we have improved our model in this paper, we run two simulations in which the satellite orbits a Dehnen model with $\gamma = 0$ and $\gamma = 1$. Fig \ref{g0.fig} shows that using the self-consistent distribution function greatly improves the accuracy of the model for the $\gamma=0$ case, and including the fast stars improves it further. There is still some discrepancy in the infall and this is most likely because the $\gamma=0$ Dehnen model has a local log-slope of the density which varies rapidly over its scale radius. In such a distribution the frictional force will always be slightly underestimated, as the locality of the density distribution assumed in equations \ref{dynfric2.eq} and \ref{dynf_full.eq} is the limiting assumption. If one wanted to more accurately reproduce the infall one would need to use a friction model that does not suffer from this assumption, such as the approach employed in \citet{ArcaSedda14}.

Fig. \ref{g1.fig} shows the cuspy case $\gamma=1$, which was already described well by the Maxwellian model in \citet{Petts15}, as $f(v\sscript{*})$ more closely resembles Maxwellian form in cuspy models. Interestingly, the model only considering the slow stars slightly over-predicts the force, and the model with all the stars reproduces the inspiral excellently. This shows that in the cuspy case the velocity dependent term in the Coulomb logarithm is small, but non-negligible. However, we would like to stress that the Maxwellian approximation, although justified in this case, performs as well as the full model by coincidence. Fig. \ref{fv.fig} shows that it over-predicts the number of slow moving stars down to $\sim0.2\unit{b}$. In general the Maxwellian approximation will not perform as well as using the self-consistent distribution function for general cuspy distributions.

In both the $\gamma=0$ and $\gamma=1$ case the stalling is very well captured by our tidal stalling mechanism. The P15 model reproduces the stalling in ${\tt gt\_D1}$ identically, but slightly under-predicts the stalling radius in ${\tt gt\_D0}$. This leads us to the conclusion that there is only one type of stalling, the tidal stalling first described in \citet{Goerdt10}. It just so happens that for distributions without a large core $b\sscript{max} \sim b\sscript{min}$ when $r\sscript{t}(r) \sim r$, which explains the success of P15 model without tidal stalling.

\subsection{Comparison with \citet{Goerdt10} and \citet{Petts15}}

\begin{figure}
 \includegraphics[width=\linewidth]{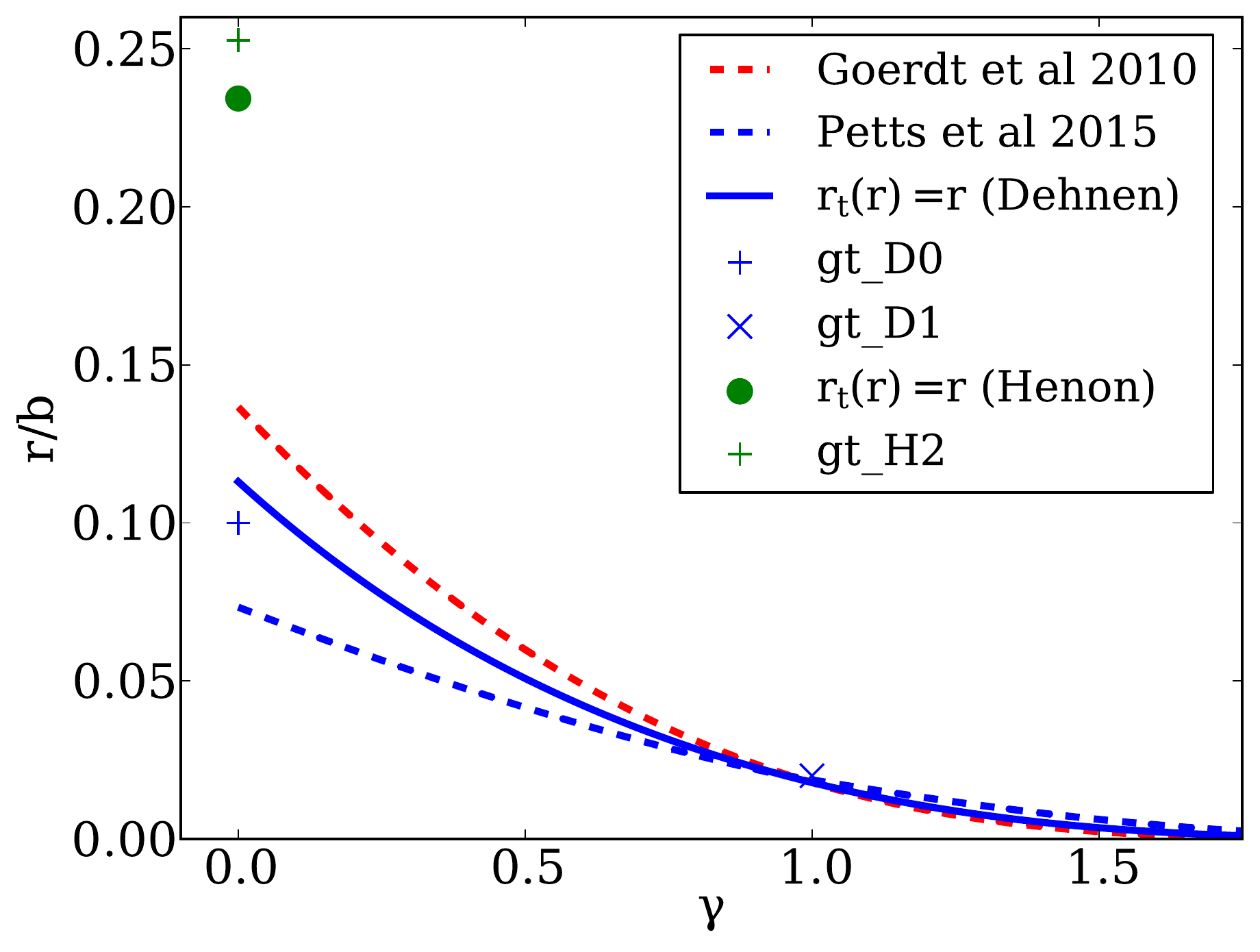}
 \caption{Stalling radii predicted by \citet{Goerdt10} (dashed red line), \citet{Petts15} (dashed blue line) and equation \ref{tidal_stalling.eq} (solid blue line) for Dehnen models as a function of $\gamma$. $N$-body results from ${\tt gt\_D0}$ and ${\tt gt\_D1}$ are marked by blue vertical and diagonal crosses respectively. The stalling radius of ${\tt gt\_H2}$ is marked by a green cross, and the estimate by equation \ref{tidal_stalling.eq} with a green circle. The \citet{Goerdt10} fit is extrapolated bellow $\gamma = 0.5$. The mass of the particle in all calculations and simulations is $3.19\times 10^{-4} \unit{M\sscript{G}}$.}
 \label{stall.fig}
\end{figure}

Fig. \ref{stall.fig} shows the stalling radii predicted by \citet{Goerdt10}, \citet{Petts15} and equation \ref{tidal_stalling.eq} for Dehnen models as a function of $\gamma$. Also displayed are the $N$-body results from simulations ${\tt gt\_D0}$, ${\tt gt\_D1}$ and ${\tt gt\_H2}$. Although all predictions agree in the cuspy regime, it is clear that equation \ref{tidal_stalling.eq} best reproduces the $\gamma = 0$ case. equation \ref{tidal_stalling.eq} also well reproduces the stalling radius in the Hénon profile (green circle on Fig. \ref{stall.fig}). A caveat is that the Goerdt prediction was fit only down to $\gamma = 0.5$, and we extrapolate their fit down to $\gamma=0$. However for galaxies with large cores, only equation \ref{tidal_stalling.eq} is good at approximating the stalling radius.

\section{discussion}
\label{discussion.ch}
Despite the successes of the analytic R06 model, it cannot be the full story. Firstly, the galactic potential is never truly harmonic in a realistic system, therefore there should be stars that do not orbit on epicycles that can contribute some frictional force. Secondly, \citet{Cole12} report an extreme example where a satellite initially inside the core actually moves outwards -- a process that they call ``dynamical buoyancy''. Such buoyancy is not captured by our semi-analytic model, nor by the R06 stalling state. 

In this section, we discuss the nature of the stalling phase. In section \S\ref{R06_general.ch}, we generalise the analytical R06 stalling state and show it is consistent with and the numerical work of \citet{Inoue09} (hereafter I09). In section \S\ref{fast_stalling.ch}, we discuss the role of the fast moving stars in the stalling phase and the related work of \citet{Inoue11} (hereafter I11). Finally, in section \S \ref{buoyancy.ch} we discuss the fast stars in the context of ``dynamical buoyancy''.

\subsection{Generalisation of the R06 model to systems with multiple satellites}

\label{R06_general.ch}

I09 performed simulations of cored dwarf galaxies containing multiple point mass globular clusters inspiraling simultaneously. The clusters perturb each others' orbits significantly throughout I09's simulations, yet the clusters still appear to stall at $M\sscript{enc} > M\sscript{s}$. I09 stated that if the co-rotating model of R06 were correct, then perturbations from other globular clusters should break the anisotropic velocity distribution found in R06, and one would expect the clusters to reach the galactic centre. In this section, we show that perturbations from other satellites are sub-dominant by generalising the R06 analysis to include multiple satellites. By starting from equation 10 in R06 and including a perturbation from $N$ other satellites, we arrive at the following equation of motion:

\begin{align}
	\begin{split}
	\bmath{\ddot{r}\sscript{p}} + \Omega^2 \bmath{r\sscript{p}} &= \frac{GM\sscript{1}(\bmath{r\sscript{1}}-\bmath{r\sscript{p}})}{|\bmath{r\sscript{1}}-\bmath{r\sscript{p}}|^3} \\
	&+ \sum_2^N \frac{GM\sscript{i}(\bmath{r\sscript{i}}-\bmath{r\sscript{p}})}{|\bmath{r\sscript{i}}-\bmath{r\sscript{p}}|^3}, \label{eqnmotion_p.eq}
	\end{split}
	\intertext{where $\bmath{r\sscript{p}}$ is the vector position of a star orbiting the combined potential of the harmonic core and system of satellites, and $M\sscript{i}$ and $\bmath{r\sscript{i}}$ are the mass and vector position of the $i^{th}$ satellite. If $M\sscript{g}(r\sscript{1}) \gg  \sum_2^N M\sscript{i}$, then:}
	\bmath{\ddot{r}\sscript{1}} + \Omega^2 \bmath{r\sscript{1}} & \simeq 0. \label{eqnmotion_s.eq}
\end{align}
Combining equations \ref{eqnmotion_p.eq} and \ref{eqnmotion_s.eq}, and substituting $\bmath{r\sscript{d}} = \bmath{r\sscript{p}}-\bmath{r\sscript{1}}$:
\begin{equation}
	\bmath{\ddot{r}\sscript{d}} + \Omega^2 \bmath{r\sscript{d}} = \frac{GM\sscript{s}\bmath{r\sscript{d}}}{|\bmath{r\sscript{d}}|^3} - \sum_2^N \frac{GM\sscript{i}(\bmath{r\sscript{p}}-\bmath{r\sscript{i}})}{|\bmath{r\sscript{p}}-\bmath{r\sscript{i}}|^3},\\
\end{equation}
\begin{equation}
	\bmath{\ddot{r}\sscript{d}} + \left(\Omega^2 + \frac{GM\sscript{1}}{|\bmath{r\sscript{1}}|^3}\right)   \bmath{r\sscript{d}} = - \sum_2^N \frac{GM\sscript{i}(\bmath{r\sscript{p}}-\bmath{r\sscript{i}})}{|\bmath{r\sscript{p}}-\bmath{r\sscript{i}}|^3}.
	\label{R06_general.eq}
\end{equation}
Hence when $|r\sscript{p}-r\sscript{i}| \gg r\sscript{d}$ the $i^{th}$ satellite is sub-dominant and solutions exist where the satellite moves in approximate epicycles around $M\sscript{1}$. For any close encounter of a star with $M\sscript{1}$ this is satisfied for all $N-1$ perturbations. It follows that only distant particles may contribute to the friction of satellite 1. However, by being distant from satellite 1, these particles are likely dominated by the potential of a different satellite, and will not interact with satellite 1 in the straight line as assumed by Chandrasekhar's formula. Therefore, the energy transfer between the distant star and satellite 1 will likely be small, if not negligible. If a satellite $M\sscript{i}$ comes close to $M\sscript{1}$, our assumptions are broken until the scattering event is complete, but after scattering $|r\sscript{p}-r\sscript{i}| \gg r\sscript{d}$ is satisfied again and $M\sscript{s}$ experiences no friction from local stars once again. This extension of the analytic R06 model is consistent with the simulations of \citet{Inoue11}.

\begin{figure}
 \includegraphics[width=\linewidth]{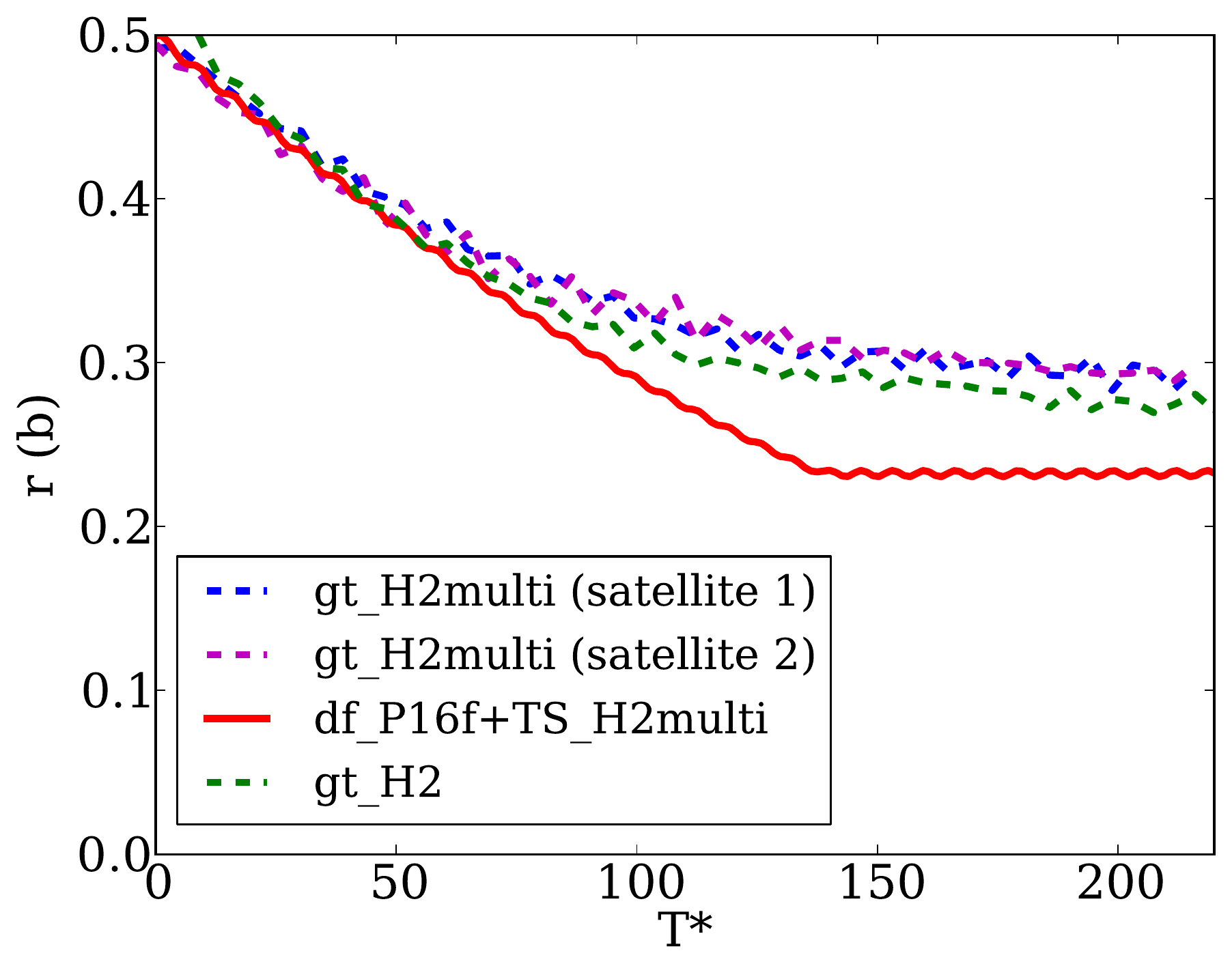}
 \caption{Evolution of the satellite orbits with time for simulations ${\tt gt\_H2multi}$ (blue  and magenta dashed lines), the late evolution of ${\tt gt\_H2}$ (green dashed line) and ${\tt df\_P16f+TS\_H2multi}$.}
 \label{multi.fig}
\end{figure}

In simulation ${\tt gt\_H2multi}$ we test the prediction of the improved analytic R06 model by setting up a fiducial case where two satellites are initially on circular orbits in the same halo. We set one satellite at $0.5b \bmath{x}$ with its velocity vector in the positive y-direction, and the other to be at $-0.5b \bmath{x}$ with its velocity vector in the z-direction. This setup ensures the satellites are maximally distant from each other during inspiral so that $|r\sscript{p}-r\sscript{i}| \gg r\sscript{d}$. From equation \ref{R06_general.eq} we predict that the stalling of each satellite should be similar to the single satellite case, as the satellites should not strongly interact. Fig. \ref{multi.fig} shows that this is the case, verifying the validity of our multi-satellite R06 formula in reproducing the results of I09.

We would like to note that for real satellites, during close encounters tidal distortions would become dominant, leading to significant distortions of the subject bodies. Satellite 1 will change shape, size and mass, but after the encounter the satellite will again find itself in a steady state with the background, as the right hand side of equation \ref{R06_general.eq} will again be negligible. Although $M\sscript{1}$ will have changed, solutions with negligible net changes of energy would still exist. The model will only fail when the satellite becomes so large compared with the core that the assumption of a point mass satellite is invalid. In this case the object will experience negligible friction in this regime regardless, as $b\sscript{min}$ will approach $b\sscript{max}$.

\subsection{Fast stars as the origin of stalling}
\label{fast_stalling.ch}

I11 showed that strongly interacting ``horn particles'' both decelerate (P-horn) and accelerate (N-horn) the satellite. In the stalling phase, \citet{Inoue11} finds that the net effect of the horn particles is a transfer of orbital energy to the satellite, opposing the frictional force from other stars. During the infall phase, however, the net effect of the P and N-horn particles is a drag on the satellite, similar to the interaction of the fast moving stars predicted by equation \ref{dynf_full.eq}.

\begin{figure}
 \includegraphics[width=\linewidth]{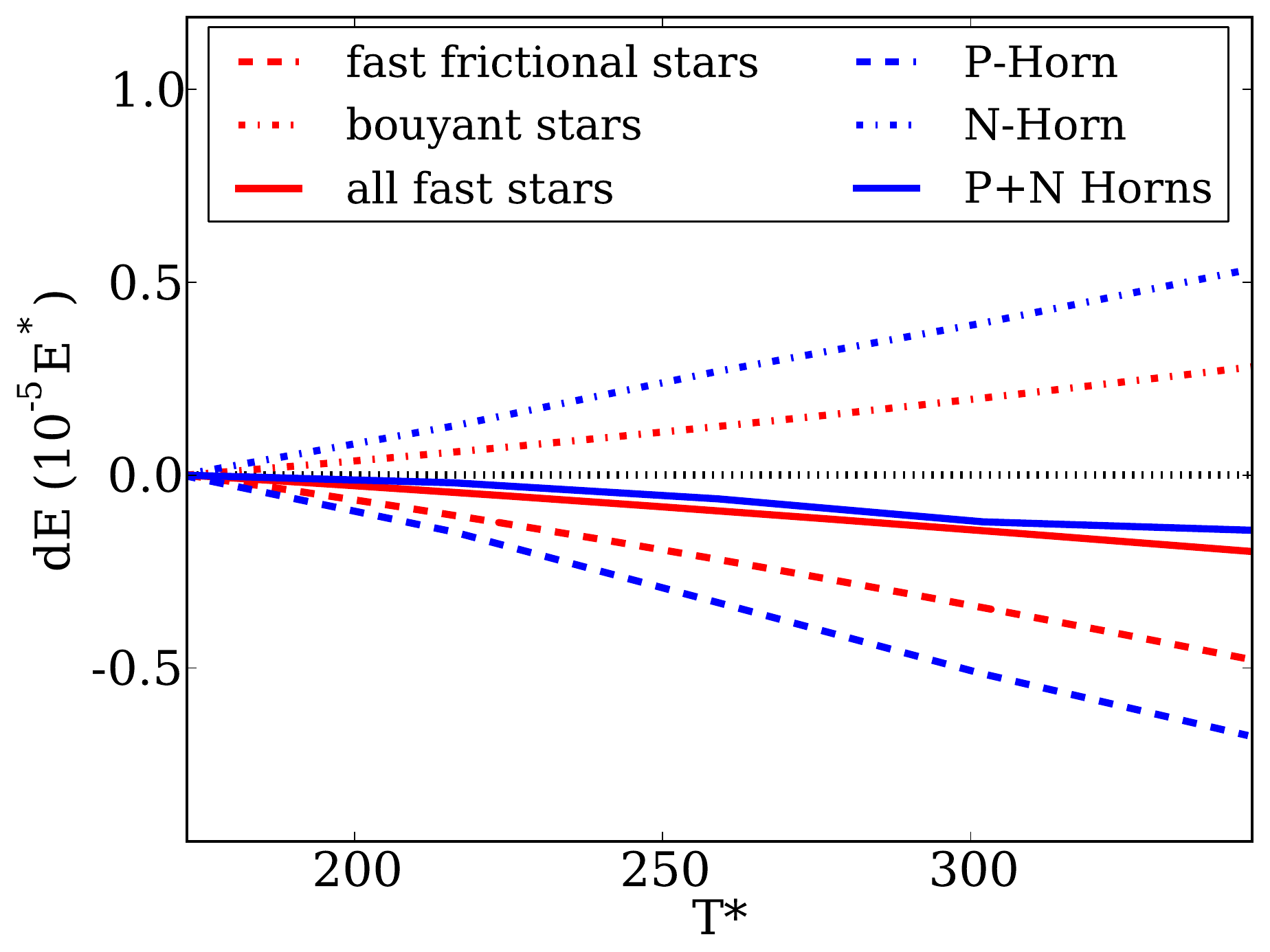}
 \caption{Cumulative energy transfer from the fast moving stars to the satellite prior to stalling in simulations ${\tt df\_P16f\_H}$ and ${\tt gt\_H}$. For ${\tt df\_P16f\_H}$ the contributions from stars that remove energy from (red dashed line) or feed energy to (red dot-dashed line) the satellite, and the net effect (solid red line) are displayed. Analogously for ${\tt gt\_H}$, the cumulative energy transfer to the satellite from the P-horn (dashed blue line), H-horn (dot-dashed blue line) and P+H horn particles (solid blue line) are shown. A dotted black line marks the x-axis for reference. Note the resemblance to fig. 6 in \citet{Inoue11}.} 
 \label{Horn.fig}
\end{figure}

Fig. \ref{Horn.fig} shows the cumulative energy transfer between fast stars and the satellite during the infall phase of simulation ${\tt df\_P16f\_H}$, as predicted by equation \ref{dynf_full.eq}. Also shown is the exchange of energy between the P and N-horn particles extracted from simulation ${\tt gt\_H}$ in the same fashion as in I11. The absolute energy transferred from each horn is larger than expected from the fast stars, however this is to be expected. The cut-off energy to define the horns is somewhat arbitrary, and one could tweak the cut-off to more closely resemble the model. However, particles artificially classified as horn particles (that didn't strongly interact with the satellite) should be equally numerous in each horn if their change in energy is instead due to two-body relaxation. Indeed, the net effect of the P and N-horn particles in ${\tt gt\_H}$ and all the fast stars in ${\tt df\_P16f\_H}$ is remarkably similar, which is strong evidence that in the inspiral phase the horn particles \textit{are} the fast moving stars. 

It is logical that the P and N-horn stars are synonymous with the fast frictional and buoyant stars in the stalling phase also. In spherical systems there is residual drag as interactions with low $V\sscript{rel}$ have a larger $b\sscript{min}$, so less of these interactions can occur. However, fig. A1 of I11 shows that during the stalling phase, the potential that stars orbit is far from spherical. It is intuitive that configurations exist where the buoyant stars can outweigh the fast frictional stars when horn-like orbits exist (those shown in fig. 10 of I11). In the stalling state, horn particles stay very close to the cluster for an extended period, allowing each interaction to occur numerous times. However, the N-horn particles transfer more energy than the P-horn particles in this state, as the strength of each individual interaction is stronger due to the $1/V^2$ dependence in equation \ref{J.eq}. We note that if all stars interact on a horn-like orbit at some point over many orbital times, this mechanism is analogous to the R06 model, whereby the time averaged contribution is zero. However, considering the effects of horn particles/fast moving stars emphasises how stalling can occur even if the potential is never truly harmonic.

\subsection{A remark on the origin of dynamical buoyancy}

\label{buoyancy.ch}

\citet{Cole12} explored different mass distributions of the Fornax dwarf spheroidal and modelled the orbital evolution of its globular clusters in a suite of 2800 $N-$body simulations. They discovered a curious effect whereby a globular cluster originating inside a constant density core is pushed outwards before stalling similarly to clusters originating further out, the authors described this as ``dynamical buoyancy''. Convergence tests ruled it out as numerical error.

The origin of this effect may owe to the increased phase space density of allowed horn orbits when the satellite is placed deep inside the pristine core. When migrating to the core from the outside, the region in which these orbits can exist expands when the satellite disrupts the core, and the satellite stalls. If the satellite originates from inside the core, it is possible that the phase-space density of horn orbits could be large enough that the buoyancy provided by N-horn stars outweighs all of the friction. In fact, since it has been verified in I11 and in section \S \ref{fast_stalling.ch} that the stalling owes to orbits of individual stars which transfer energy into the satellite, one \textit{must} be able to construct a fiducial system whereby the N-horn particles dominate over the frictional particles. This setup will of course be unstable and the net effect is ``dynamical buoyancy'' as discussed in \citet{Cole12}. The satellite orbit would stop increasing when the density of these orbits decreases significantly enough that the net force on the satellite is zero. Our current model cannot capture this effect since we do not explicitly model the effect of N-horn versus P-horn interactions. We leave such a study of dynamical buoyancy to future work.

\section{conclusions}
\label{conclusion.ch}

We have shown that Chandrasekhar's dynamical friction model considering only two-body encounters is sufficient to explain the inspiral of satellites in constant density profiles, so long as one uses the self-consistent distribution function of velocities instead of the usually assumed Maxwellian distribution. In particular, we show that we can reproduce the ``super-Chandrasekhar'' phase, suggesting that it does not owe to resonance. The Chandrasekhar formula probably works so well because the potential is never truly harmonic in any physically reasonable distribution. The agreement is improved further by including the usually neglected contribution of the fast moving stars, which contributes a non-negligible portion of the drag inside the core.

However, even after including the correct background distribution function and the effect of fast moving stars, we find that we are not able to reproduce the stalling observed in large constant density cores such as in the Hénon Isochrone Model. Following \citet{Goerdt10}, we show that such large-core stalling occurs in the same manner as it does for cusps. The infalling satellite tidally disrupts the core when $r\sscript{t}(r\sscript{a}) = r\sscript{a}$. For cusped background models this occurs at $M\sscript{s} \sim M\sscript{enc}$. However, for cored backgrounds, the satellite tidal radius can become very large indeed. This leads to stalling at many times the radius at which the mass in background stars approaches the satellite mass. In our model, $f(v\sscript{*})$ is derived from the distribution function of the background density distribution and our model has no free or tuned parameters. As such, it should be general for any model with a cored or cusped centre. 

Finally, we suggest that the dynamical friction core-stalling can be understood in two different ways. For a perfectly harmonic background with a single point mass satellite, R06 demonstrated that there exist stable solutions with no net momentum exchange between the satellite and the background. We generalised this model in section \ref{discussion.ch}, showing that the same should be true when multiple satellites are present. While the satellite and the background likely reach an approximation to this state, the correspondence cannot be perfect. Secondly, the ``dynamical buoyancy'' effect discussed in \citet{Cole12} is not captured by our semi-analytic model, nor by the R06 stalling state. Instead, the answer may lie in the frictional force coming from stars moving faster than the satellite. \citet{Inoue11} showed that strongly interacting ``Horn'' stars can both decelerate (P-horn) and accelerate (N-horn) the satellite depending on their relative orbital phase. For a large-cored background, the cumulative effects of P and N-horn stars approximately cancel the friction experienced in the core region, leading to core-stalling as in the R06 model. However, configurations can exist where it is possible for N-horn stars to dominate over P-horn ones, if a satellite begins its life deep inside the constant density core. This is a possible explanation for \citet{Cole12}'s dynamical buoyancy; however, a full proof will require further investigation beyond the scope of this work.

\section*{Acknowledgements}
We would like to thank the anonymous referee for very helpful comments that greatly improved the scientific content of this paper.
JAP would like to thank the University of Surrey for the computational resources required to perform the high resolution $N$-body simulations. JIR would like to acknowledge support from STFC consolidated grant ST/M000990/1 and the MERAC foundation. JAP would like to thank Michelle L. M. Collins for te$\chi$nical support.

\bibliography{jpetts2016}
\bibliographystyle{mn2e,natbib}

\label{lastpage}

\end{document}